\documentclass[preprint,12pt]{elsarticle}
\usepackage{pdflscape}
\usepackage{threeparttable}
\usepackage{threeparttablex} 
\usepackage{amssymb}
\usepackage{amsmath}
\usepackage{booktabs}
\usepackage{tabularx}
\usepackage{multirow}
\usepackage{listings}
\usepackage{xcolor}
\usepackage{hyperref}
\definecolor{codegray}{gray}{0.95}
\lstdefinestyle{cli}{
  backgroundcolor=\color{codegray},
  basicstyle=\ttfamily\small,
  breaklines=true,
  frame=none,
  columns=fullflexible,
  showstringspaces=false,
  xleftmargin=1em,
  framexleftmargin=1em,
}
\newcounter{bla}

\usepackage{gensymb}
\usepackage{xspace}
\usepackage{siunitx}
\newcommand{\ii}{\mathrm{i}}
\newcommand{\dd}{\mathrm{d}}
\newcommand{\defeq}{\overset{\mathrm{def}}{=}}
\newcommand{\adda}{\texttt{ADDA}\xspace}
\newcommand{\ddscat}{\texttt{DDSCAT}\xspace} 
\newcommand{\ifdda}{\texttt{IFDDA}\xspace}
\newcommand{\hb}[1]{\hat{\mathbf{#1}}}
\newcommand{\cmd}[1]{\texttt{\detokenize{#1}}}
\makeatletter
\def\ps@pprintTitle{%
  \let\@oddhead\@empty
  \let\@evenhead\@empty
  \def\@oddfoot{\reset@font\hfill\thepage\hfill}
  \let\@evenfoot\@oddfoot}
\makeatother
\begin{document}
\sloppy
\begin{frontmatter}
\title{Floating-point--consistent cross-verification methodology for reproducible and interoperable DDA solvers with fair benchmarking}
\author[1]{Clément Argentin\corref{author}}
\author[2]{Patrick C. Chaumet}
\author[3]{Michel Gross}
\author[1]{Maxim A. Yurkin}
\cortext[author] {Corresponding author.\\\textit{E-mail address:} yurkin@gmail.com}
\address[1]{Université Rouen Normandie, INSA Rouen Normandie, CNRS, CORIA UMR 6614, Rouen, 76000, France}
\address[2]{Institut Fresnel, Aix Marseille Univ, CNRS, Centrale Marseille, Marseille, 13013, France}
\address[3]{Université Montpellier II, CNRS, L2C UMR 5221, Montpellier, 34090, France}
\begin{abstract}
The discrete dipole approximation (DDA) is a widely used and versatile numerical method for solving electromagnetic scattering by arbitrarily shaped objects. Despite its popularity, quantitative comparisons between independent implementations remain challenging due to differences in linear-system conventions, solver settings, and default numerical parameters. In this work, we introduce a unified software-assisted methodology for cross-verification and benchmarking of three major open-source DDA solvers: \ddscat, \adda, and \ifdda. We demonstrate how machine-precision agreement can be achieved across implementations by aligning all free parameters and provide practical equivalence tables enabling reproducible and interoperable simulations. Using this methodology, we perform systematic CPU and GPU performance comparisons covering OpenMP, MPI, and CUDA/OpenCL parallelization. Beyond benchmarking, our approach serves as a practical guide for configuring consistent DDA simulations and for understanding how precision, solver choice, and hardware architecture affect runtime, scalability, and accuracy in computational light-scattering studies. The software package also supports regression testing and bitwise reproducibility verification for future code releases.

\noindent \textbf{PROGRAM SUMMARY}

\begin{small}
	\noindent
	{\em Program Title:} dda-bench                                          \\
	{\em CPC Library link to program files:} \\
	{\em Developer's repository link:} \url{https://doi.org/10.5281/zenodo.18836855} \\
	{\em Licensing provisions(please choose one):} GPLv3  \\
	{\em Programming language:} Python                                 \\
	{\em Nature of problem:}\\
	Independent implementations of the discrete dipole approximation (DDA), such as \ddscat, \adda, and \ifdda, are widely used for electromagnetic scattering by arbitrarily shaped objects. However, quantitative cross-code comparisons and fair performance benchmarks remain difficult because the codes may differ not only in linear-system conventions, but also in default numerical parameters, solver settings, polarizability and interaction formulations, and output conventions or units. Small mismatches can modify iterative convergence histories and lead to discrepancies that mask genuine algorithmic or hardware effects. Users and developers therefore lack a practical and reproducible way to (i) verify that two codes are solving the same numerical problem, (ii) assess the number of matching digits that should be expected when specific parameters differ, and (iii) benchmark runtimes without confounding accuracy differences.\\
	{\em Solution method:}\\
	The presented software provides a lightweight, command-line--based framework to compare numerical DDA implementations through their executables. The user specifies, in a dedicated input file, the command-lines to be tested for each code or code version. The Python wrapper automatically executes the specified commands, collects the generated output files, and extracts the physical quantities selected by the user for comparison. The extracted data are converted automatically into consistent definitions and units when necessary. Agreement between simulations is quantified using a matching-digits metric, which directly reflects floating-point consistency. This approach enables straightforward comparison between different DDA codes or between multiple versions of the same solver, thus supporting regression testing, cross-verification, and fair benchmarking once equivalent configurations are enforced.\\
	{\em Additional comments including restrictions and unusual features:}\\
	The software is limited to DDA implementations that provide a command-line interface and assumes that the user has already installed and compiled the corresponding codes. The framework itself acts as a Python wrapper that operates on the generated executables. This design choice facilitates the practical use of the framework as a lightweight Python package that can be installed and imported easily, while allowing users to track and compare different releases of the same code and to freely choose which executable versions are used in a given comparison.\\
\end{small}
\end{abstract}
\end{frontmatter}
\section{Introduction}
\label{sec:intro}

Light scattering by particles is a fundamental process governing a wide range of optical phenomena~\cite{mishchenko_electromagnetic_2014}. Quantitative modeling of scattering, absorption, and near-field effects is essential for interpreting experiments and designing optical systems. Analytical solutions of Maxwell's equations are limited to a few idealized geometries (e.g., spheres or spheroids), making numerical approaches indispensable for realistic systems of arbitrary shape, composition, and environment.

Among general numerical techniques, several broad families of methods exist: finite-difference time-domain (FDTD)~\cite{taflove2005computational}, finite-element (FEM)~\cite{monk2003finite}, boundary-element or surface-integral equation (BEM/SIE)~\cite{harrington1993field}, and volume-integral equation (VIE)~\cite{yurkin2018VIE} ones. Within the latter category, the discrete dipole approximation (DDA) is commonly used due to its conceptual simplicity, modularity, and favorable scaling with particle size and wavelength~\cite{chaumet_review_2022, yurkin_review_2023}. Introduced in Ref.~\cite{purcell_scattering_1973}, and further developed in Refs.~\cite{draine_discrete_1988, draine_DDSCAT_1994}, DDA discretizes the scatterer into an array of polarizable voxels (dipoles) and computes their induced polarizations by solving a large linear system derived from the volume-integral form of the Maxwell's equations. Once the voxel polarizations are known, all scattering observables, including cross sections, Mueller matrices, and field distributions, can be computed in both near- and far-field regions. DDA is employed for arbitrarily shaped particles, and complex optical environments, across disciplines from astrophysics~\cite{draine_discrete_1988} and atmospheric science~\cite{sorensen2018light} to nanotechnology~\cite{amendola2016surface} and biomedical applications~\cite{Yurkin2005bio}. Moreover, as it is a ``numerically exact'' method, DDA is often used as a reference to validate or correct approximate analytical models~\cite{yon2008extension, argentin2023electromagnetic}.

Over the past three decades, several open-source DDA codes have been developed, each optimized for different architectures and applications. The most widely used are \ddscat~\cite{draine_DDSCAT_1994},
\adda~\cite{yurkin_ADDA_2011}, and \ifdda~\cite{chaumet_ifdda_2021}.
An extensive list of publicly available DDA implementations and related tools is available online~\cite{ADDA_GitHub_Wiki,Wikipedia}.

Despite their shared theoretical foundation, comparing the numerical performance and accuracy of different DDA implementations remains nontrivial. Discrepancies often arise from subtle differences in polarizability formulations, iterative solvers, or linear system conventions. As a result, cross-code benchmarking~\cite{penttila_comparison_2007} is challenging to interpret, since the codes have different accuracy and different computational performance.

The aim of this paper is therefore twofold. 
First, we introduce a unified software-assisted methodology for floating-point--consistent cross-verification between independent DDA implementations. By harmonizing all free and hidden parameters, we demonstrate that solvers can agree up to machine precision in double precision (i.e.\ 14--15 correct digits). Focusing on \ddscat, \adda, and \ifdda, we provide practical equivalence tables that enable reproducible and interoperable DDA simulations. Because these implementations rely on different linear-system conventions, and FFT backends, strict bitwise reproducibility is not achievable across codes, as even tiny differences in floating-point operation ordering lead to round-off divergence. Such limitations are well-known concerns in high-performance scientific computing, particularly for parallel numerical simulations~\cite{wiesenberger2019reproducibility,COLLANGE201583}.

Second, using this consistent numerical baseline, we conduct a fair benchmarking study across CPUs and GPUs, covering OpenMP and MPI parallelization on CPUs and CUDA/OpenCL acceleration on GPUs. Performance differences observed in this study therefore reflect algorithmic and hardware architectural effects rather than configuration inconsistencies.

Beyond cross-verification, the software package also supports comparisons between different versions of the same implementation. In that regime, true bitwise reproducibility is achievable, enabling continuous integration and continuous delivery (CI/CD) regression tests and version-to-version consistency checks. For instance, this will complement the automatic tests already present in the \adda workflow.

Finally, this paper aims to serve as a reference for best practices in setting up DDA simulations, and understanding the impact of implementation choices. Section~\ref{sec:overview} provides a comparative overview of the three DDA codes and their internal features; Section~\ref{sec:interoperability} acts as a methodological guide and a technical prerequisite for the performance benchmarks presented in Sections~\ref{sec:comparison} and~\ref{sec:comparison_gpu}.

\section{DDA implementations and workflows}
\label{sec:overview}
The main objective of this section is to provide a concise and self-consistent description of the DDA workflow, introducing its key internal components to facilitate the discussion in later sections. We also summarize distinctive features of each considered implementation so that this paper serves as a practical comparative overview of the three main open-source DDA codes.

\begin{table}[!ht]
\centering
\tiny
\begin{threeparttable}
\caption{Overview of selected features in \ddscat, \adda, and \ifdda.\tnote{a}}
\label{tab:features_comparison}
\begin{tabularx}{\textwidth}{lXXX}
\toprule
Feature / Code & \ddscat & \adda & \ifdda \\
\midrule
Application focus & general & general & multi-layered substrates and microscopy \\
Community and contributions & open source (website) & open source (GitHub) & open source (GitLab) \\
User interface & CLI-oriented (configuration file) with optional GUI & CLI-oriented with optional GUI & GUI-oriented with optional CLI \\
Language & Fortran~90 & C99 & Fortran~77 \\
Hardware support & CPU only & CPU and GPU (OpenCL) & CPU and GPU (CUDA) \\
Parallelization & OpenMP + MPI (partial)\tnote{b} & MPI (full) & OpenMP \\
Orientation averaging & Simpson rule & Romberg integration & -- \\
Precision & single (default) or double & double (default) & double (default) or single \\
Incident wave types & plane wave only & plane wave and others & plane wave and others \\
FFT algorithms & GPFA, Intel~MKL & GPFA, FFTW & Singleton, FFTW \\
DDA formulations & POI, FCD & POI, FCD, IGT & POI, FCD, IGT \\
Polarizability models & LDR, CLDR, FCD & CM, RR, LDR, CLDR, FCD, LAK, DGF & CM, RR, LDR, FCD, LAK, GB \\
Initial field & zero only & zero and others & zero and others \\
Preconditioner & left (Jacobi) or none & none & left, right, both, (Chan) or none \\
Volume correction & yes & optional (yes/no) & no \\
Iterative solver & BCGS2, BiCGStab, GPBiCG, QMR, PETRKP & BCGS2, BiCG, BiCGStab, CGNR, CSYM, QMR, QMR2 & GPBiCG, GPBiCGStab, BiCGStab, CG, QMR, IDRS and many others \\
Unique features & periodic target & Bessel beams, EELS/CL, sparse mode, cuboid voxels, homogeneous substrate (with 3D FFT) & multi-layered substrates, microscopy, fully anisotropic target\\
\bottomrule
\end{tabularx}
\begin{tablenotes}
\footnotesize
\item[a] Used abbreviations are explained in the text.
\item[b] Only for different particle orientations (orientation averaging) in contrast to distributed-memory parallelization of a single simulation in \adda.
\end{tablenotes}
\end{threeparttable}
\end{table}

All three implementations: \ddscat, \adda, and \ifdda, aim to simulate the interaction of electromagnetic waves with arbitrarily shaped particles or objects. Both \ddscat and \adda are primarily designed for general light-scattering problems, and particularly suited for databases and large-scale parametric studies. In contrast, \ifdda focuses on applications involving complex optical setups, including multilayered substrates and microscopy configurations. It can generate observable quantities such as intensity images making it especially suitable for modeling and visualization tasks in applied or experimental optics~\cite{durdevic2022biomass}.

All three codes implement similar DDA workflows but differ in numerical details and hardware optimization strategies. Each supports several Krylov-space iterative solvers (of the conjugate-gradient, CG, family), which are used for solving the large sparse linear systems. As noted in Ref.~\cite{chaumet_solver_2024}, it is generally not possible to identify a single solver that performs best in all scenarios, since performance depends strongly on particle morphology and refractive index. In Ref.~\cite{chaumet_solver_2024}, the QMR solver (used in \ifdda) is implemented in its non-symmetrized form (as in \ddscat); when symmetrized, as done in \adda, the number of matrix–vector products is reduced by a factor of two, resulting in similar convergence rates and execution times as other CG-family solvers. Consequently, solver performance comparisons between codes must take these implementation details into account. For consistency, all performance results presented in Secs.~\ref{sec:comparison} and~\ref{sec:comparison_gpu} are therefore based on the BiCGStab solver.

DDA has undergone several refinements in polarizability models and interaction formulations~\cite{yurkin_review_2007,yurkin_review_2023}. The classical approach, i.e., point-dipole interaction (POI) combined with the lattice dispersion relation (LDR) polarizability~\cite{draine_DDSCAT_1994}, remains widely used, while more modern formulations such as the filtered coupled dipoles (FCD)~\cite{piller1998increasing,yurkin_fcd_2010} and the integrated Green's tensor (IGT)~\cite{Chaumet2004IGT,SMUNEV2015Rect} have been developed to improve both accuracy and iterative convergence~\cite{yurkin_review_2023}. Fortunately, all three codes support at least one of these advanced formulations. FCD and IGT come with their own polarizability prescriptions, while POI support many others~\cite{yurkin_review_2007}: Clausius-Mossotti (CM), radiative-reaction (RR) correction, digitized Green's function (DGF / GB), Lakhtakia (LAK), and lattice dispersion relation (LDR) or corrected one (CLDR).

The incident-field models differ across implementations. \ddscat primarily supports plane-wave illumination. \adda additionally provides (approximate) Gaussian beams and point-dipole excitation, accepts user-specified arbitrary fields from file, and in its master branch provides a variety of built-in Bessel beams~\cite{Glukhova2022Bessel}. Development branches further implement electron energy-loss spectroscopy (EELS) and cathodoluminescence (CL) for particles in arbitrary host medium~\cite{Kichigin2023EELS}. \ifdda supports physical Gaussian beams and a wider set of structured illuminations (circularly polarized, multi-plane-wave, speckle, confocal, and arbitrary imported fields), and allows targets to be embedded in multilayer substrates. It can also generate microscopy images (e.g., bright-field, dark-field, phase-contrast, holography) from the computed fields.

Iterative convergence may be accelerated by advanced initial fields and appropriate preconditioners. \ifdda implements several advanced initial guesses such as uGu that can be combined with Chan-type preconditioning~\cite{Chan1988Circulant} to reduce iteration counts~\cite{CHAUMET2023comparative}. Recently, an implementation of a three-level circulant preconditioner extended this work and has been shown to further reduce iteration counts~\cite{LANIER2026109741}. \adda provides, in addition to the classic zero and incident field, a Wentzel–Kramers–Brillouin (WKB) initial field option, while \ddscat provides the zero-field initialization with a possible left Jacobi preconditioning for BiCGStab solver. It should be noted that although these initial fields may be beneficial in some cases, they require initialisation time to be calculated and can therefore increase calculation time if used outside their limits. A recent study~\cite{argentin:inital_guesses} compared the initial WKB field with the uGu formulation~\cite{Chaumet_uGu_22} and showed that their performance was broadly equivalent. Beyond rigorous but computationally expensive DDA, \ifdda also offers approximate direct methods that are effective for multilayer transmission-reflection problems~\cite{Chaumet2025MLB}.

All implementations are portable across operating systems and require few external dependencies. By default, \ddscat uses the self-contained GPFA FFT algorithm~\cite{temperton_gpfa_1992}, which requires only a Fortran compiler (e.g., \texttt{gfortran}), making it highly portable. \adda can also be compiled with GPFA (named Temperton), ensuring compatibility without external libraries (except for \texttt{gcc} and \texttt{gfortran} compilers). However, as shown in Sec.~\ref{sec:cpu_times}, GPFA underperforms compared to optimized FFT libraries such as Intel~MKL routine DFTI and FFTW~\cite{frigo1998fftw}. The same applies to \ifdda with the Singleton algorithm~\cite{Singleton1967FFT}. Therefore, the recommended configurations are \ddscat MKL, \adda FFTW, and \ifdda FFTW. GPU modes introduce further dependencies: \adda requires OpenCL and clFFT, while \ifdda relies on CUDA. \adda also supports package managers such as \texttt{Spack}~\cite{gamblin2015spack} or \texttt{Nix}, facilitating portable and reproducible installations in contained environments and high-performance computing (HPC) clusters.

HPC has strongly influenced the evolution of the DDA codes. \ddscat and \ifdda are optimized for shared-memory architectures using OpenMP, whereas \adda employs MPI to leverage distributed-memory clusters. Recently, GPU accelerators have become standard for accelerating matrix–vector operations. \ifdda uses CUDA, the most mature and widely optimized GPU framework, but limited to NVIDIA hardware, while \adda uses OpenCL to ensure cross-vendor portability across NVIDIA, AMD, and Intel GPUs, at the cost of lower performance.

Floating-point precision is another important design choice. Double precision ensures better numerical stability but increases memory traffic and runtime, whereas single precision offers higher arithmetic throughput and lower memory requirements, especially beneficial for GPUs, where available memory is often limited. As demonstrated in Secs.~\ref{sec:cpu_single} and~\ref{sec:precision_effects}, single precision can yield substantial speedups with acceptable error levels for well-conditioned problems. However, convergence of the iterative solver may degrade for some cases, illustrating the trade-offs between performance, portability, and numerical accuracy inherent to the modern DDA implementations.

We used the latest releases (or commits) of each code: \ddscat~v7.3.4, \ifdda~v1.0.27, and \adda~v.1.5.0-alpha3 (commit \texttt{b03d648}). Minor, non-intrusive modifications were applied to enable machine-precision cross-comparisons in our benchmark tests; all such changes are documented in~\ref{Appendix:Modifications}. Build configurations (compilers, flags, libraries) are reported alongside the benchmarks to ensure reproducibility.

\section{Accuracy, interoperability, and guidelines}
\label{sec:interoperability}

This section serves as both a methodological guide and a technical prerequisite for the performance benchmarks. We first analyze the main sources of numerical error in DDA simulations and clarify what ``accuracy'' means in this context. We then outline the interoperability methodology used to align the parameters of different DDA codes, providing practical guidance for achieving consistent, reproducible, and high-precision results.

\subsection{Understanding accuracy in DDA simulations}
\label{subsec:accuracy_dda}

The accuracy of a DDA simulation depends on several interacting factors. DDA is fully deterministic, and its errors can therefore be decomposed into well-defined categories. Understanding these sources of error is essential for interpreting the observed differences between simulation results and for assessing the level of agreement achievable between different implementations.

\begin{itemize}
    \item \textbf{Iterative convergence error.}  
    The iterative solver used to obtain the internal electric field must reach a specified tolerance $\eta$, defined as the relative norm of the residual $\lVert \mathbf{r} \rVert / \lVert \mathbf{b} \rVert$. The corresponding relative error in the field and derived quantities (e.g., extinction cross section) is typically $\eta$, which is generally negligible in comparison to the method error below (for the used default values of $\eta=10^{-4}$ or $10^{-5}$). Still, it may become an issue if $\eta$ is increased in an effort to accelerate computations~\cite{Kichigin2023EELS}.
    
    \item \textbf{Method error.}  
    This represents the intrinsic accuracy of DDA compared to the exact electromagnetic solution. It is composed of two main contributions:
    \begin{itemize}
        \item \emph{Discretization error} which appears from approximating the target by a finite number of voxels $N$. Accuracy improves with increasing discretization grid, e.g., the number of voxels along the $x$-axis $n_x$, or equivalently, with smaller voxel spacing $d$~\cite{Yurkin2006Convergence}.
        \item \emph{Shape error} due to mismatch between the discretized geometry and the actual particle shape. Apart from grid refinement, this error may be decreased by the weighted discretization~\cite{piller1997influence}.
    \end{itemize}
    The required or expected method error largely depends on application. Typically, 1\% error is considered satisfactory for integral scattering quantities, while up to 10\% can be fine for angle-resolved one. We stress, however, that better accuracy is perfectly possible, and relative errors down to $10^{-7}$ have been demonstrated~\cite{yurkin_light_2013}.
    \item \textbf{Round-off error.}  
    Finite-precision arithmetic leads to accumulation of rounding errors over many iterations. When the number of matrix--vector products and iteration count $N_{\mathrm{it}}$ are large, these small numerical fluctuations (about $10^{-7}$ or $10^{-16}$ in single and double precision, respectively) accumulate and ultimately limit the number of matching digits to that set by the convergence threshold $\eta$.
\end{itemize}
In practice, when comparing different implementations or numerical setups, two typical levels of agreement can be identified:
\begin{itemize}
    \item \textbf{Machine-precision agreement.}  
    When all simulation parameters are aligned across codes, one can achieve near machine-level agreement, typically 10--15 matching digits in double precision. This defines the upper limit of reproducibility for deterministic DDA solvers.

    \item \textbf{$\boldsymbol{\eta}$-level matching.}  
    When comparing results across codes with identical physical settings but different ways to solve the resulting linear system, the agreement is limited by the latter.
\end{itemize}

Understanding these different sources of error provides a foundation for the cross-code verification methodology described in the following subsection. In practice, machine-level agreement is only achievable when the same free parameters are used (see Sec.~\ref{sec:default_parameters}), identical precision modes are selected, and linear system conventions (see Sec.~\ref{sec:linear_systems}) agree. When any of these conditions differ, one can only hope to reach the $\eta$-level agreement (see Sec.~\ref{sec:free_parameters}), but in some cases only agreement within the method error is obtained. Note also, that $\eta$-level agreement is sufficient for most physical applications and can be considered sufficiently good for software testing, since $\eta$ can be varied in a wide dynamic range.

\subsection{Different linear systems (\adda, \ifdda, \ddscat)}
\label{sec:linear_systems}
In this section, we aim to identify the conditions under which the linear systems used in different DDA implementations are not just equivalent (which is always the case), but lead to exactly the same convergence trajectory of standard iterative methods. This marks the difference between $\eta$-level and machine-precision agreement in the computed quantities.

In the following, we express all quantities using the SI units.
All permittivities and refractive indices are expressed relative to the vacuum background. The relative permittivity tensor is defined as
\begin{equation}
\overline{\boldsymbol{\varepsilon}}(\mathbf{r}) \defeq
\begin{cases}
\overline{\mathbf{I}}, & \mathbf{r} \in V_\mathrm{ext}, \\
\overline{\boldsymbol{\varepsilon}}_\mathrm{p}(\mathbf{r})/\varepsilon_0, & \mathbf{r} \in V_\mathrm{int},
\end{cases}
\end{equation}
where $\varepsilon_0$ is the vacuum permittivity and $\overline{\boldsymbol{\varepsilon}}_\mathrm{p}(\mathbf{r})$ is the permittivity of the particle. The complex relative refractive index and electric susceptibility are defined as
\[
\overline{\mathbf{m}}(\mathbf{r}) = \sqrt{\overline{\boldsymbol{\varepsilon}}(\mathbf{r})}, 
\quad 
\overline{\boldsymbol{\chi}}(\mathbf{r}) \defeq \overline{\boldsymbol{\varepsilon}}(\mathbf{r}) - \overline{\mathbf{I}}.
\]
The wavenumber in vacuum is defined as $k = \omega \sqrt{\varepsilon_0 \mu_0}$, where $\mu_0$ is the vacuum permeability. The free-space dyadic Green's function is given by:
\begin{equation}
\overline{\mathbf{G}}(\mathbf{r}, \mathbf{r}')
=
\frac{\exp(\ii kR)}{4\pi R}
\left[
\left( \overline{\mathbf{I}} - \frac{\mathbf{R} \otimes \mathbf{R}}{R^2} \right)
+
\frac{\ii kR - 1}{k^2 R^2}
\left(
\overline{\mathbf{I}} - 3 \frac{\mathbf{R} \otimes \mathbf{R}}{R^2}
\right)
\right],
\label{Eq:Green_tensor}
\end{equation}
with $\mathbf{R} = \mathbf{r} - \mathbf{r}'$, $R = \|\mathbf{R}\|$, and the dyad $\mathbf{R} \otimes \mathbf{R}$.

When referring to code-specific linear systems, we use superscripts to distinguish between systems: $\mathbf{A}^p$ for the standard polarization form~\cite{draine_discrete_1988} (\ddscat~\cite{draine_DDSCAT_1994}), $\mathbf{A}^x$ for the symmetrized form~\cite{yurkin_review_2023} (\adda~\cite{yurkin_ADDA_2011}), and $\mathbf{A}^E$ for the internal field form~\cite{chaumet_review_2022} (\ifdda~\cite{chaumet_ifdda_2021}).

\subsubsection{Standard DDA (polarization form, e.g., \ddscat)}
In SI units, the linear system solved in the standard polarization formulation~\cite{yurkin_review_2023}, after discretization, is:
\begin{equation}
\overline{\boldsymbol{\alpha}}_i^{-1} \mathbf{p}_i - k^2 \sum_{j \neq i} \overline{\mathbf{G}}_{ij} \mathbf{p}_j = \mathbf{E}_i^\mathrm{inc}
\label{Eq:standard_DDA}
\end{equation}
where $\overline{\mathbf{G}}$ is the free-space dyadic Green's tensor defined in Eq.~\eqref{Eq:Green_tensor}, and $\overline{\boldsymbol{\alpha}}$ is the polarizability tensor. Note that the voxel moments $\mathbf{p}_i$ and polarizabilities $\overline{\boldsymbol{\alpha}}_i$ are scaled by a factor of $1/\varepsilon_0$ (see Eq.~(10) in Ref.~\cite{yurkin_review_2023}) compared to their conventional SI definitions to ensure consistency with the linear form in the CSG convention. The latter is different from Eq.~\eqref{Eq:standard_DDA} only by the lack of $k^2$ factor (see Eq.~(20) in Ref.~\cite{yurkin_review_2007}).

This can be written in compact matrix form as:
\begin{equation}
\mathbf{A}^p \mathbf{p} = \mathbf{E}^\mathrm{inc}
\end{equation}
where the matrix elements $\mathbf{A}^p_{ij}$ are defined as:
\begin{equation}
\mathbf{A}^p_{ij} = 
\begin{cases}
\overline{\boldsymbol{\alpha}}_i^{-1}, & i = j \\
-k^2 \overline{\mathbf{G}}_{ij}, & i \neq j
\end{cases}
\end{equation}
This matrix $\mathbf{A}^p$ is complex symmetric for arbitrary complex symmetric $\overline{\boldsymbol{\alpha}}_i$. 

\subsubsection{Symmetrized-form (change of variable, e.g., \adda)}
Change of variable based on the square root of the polarizability tensor~\cite{yurkin_review_2023, erratum_yurkin_review_2023}:
\begin{equation}
\mathbf{b}_i = \overline{\boldsymbol{\beta}}_i \mathbf{E}_i^\mathrm{inc}, 
\quad 
\mathbf{x}_i = \overline{\boldsymbol{\beta}}_i^{-1} \mathbf{p}_i
\end{equation}
where $\overline{\boldsymbol{\beta}}_i = \overline{\boldsymbol{\alpha}}_i^{1/2}$. Replacing into the standard DDA form leads to a transformed system:
\begin{equation}
\mathbf{A}^x \mathbf{x} = \mathbf{b}
\end{equation}
with matrix elements given by:
\begin{equation}
\mathbf{A}^x_{ij} =
\begin{cases}
\mathbf{I}, & i = j \\
-k^2\overline{\boldsymbol{\beta}}_i \, \overline{\mathbf{G}}_{ij} \, \overline{\boldsymbol{\beta}}_j, & i \neq j
\end{cases}
\end{equation}
This matrix $\mathbf{A}^x$ is complex symmetric for arbitrary complex symmetric $\overline{\boldsymbol{\alpha}}_i$ and has unit diagonal (i.e, it is Jacobi-preconditioned). 

\subsubsection{Internal field form (e.g., \ifdda)}
The local internal electric field satisfies in SI units:
\begin{equation}
    \mathbf{E}_i^\ell = \mathbf{E}^\mathrm{inc}_i + \sum_{j \neq i} k^2\overline{\mathbf{G}}_{ij} \, \overline{\boldsymbol{\alpha}}_j \, \mathbf{E}_j^\ell
\end{equation}
The structure of the linear system is different by a factor of $k^2$ compared to Ref.~\cite{chaumet_review_2022}-CGS form, due to the transformation of $\overline{\mathbf{G}}$. The polarizabilities $\overline{\boldsymbol{\alpha}}_i$ are scaled by a factor $1/\varepsilon_0$ relative to their strict SI definitions (as discussed previously). Additionally, while in Ref.~\cite{chaumet_review_2022} the susceptibility is defined as $\overline{\boldsymbol{\chi}} = \frac{1}{4\pi}(\overline{\boldsymbol{\varepsilon}}_r - \mathbf{I})$ and uses a Green's tensor without the $1/4\pi$ prefactor, we instead adopt the standard SI definition $\overline{\boldsymbol{\chi}} = \overline{\boldsymbol{\varepsilon}}_r - \mathbf{I}$ and include the $1/4\pi$ factor directly in our Green's tensor (see Eq.~\eqref{Eq:Green_tensor}).

Here $\mathbf{E}_i^\ell$ is the local field, also called microscopic field~\cite{jackson_classical_1998} (according to \ddscat, see Sec.~25 of the manual~\cite{draine2013userguidediscretedipole}), or excited field~\cite{yurkin_review_2007} $\mathbf{E}^\mathrm{exc}$, and corresponds to the total field $\mathbf{E}_i$ without the field induced by the voxel on itself. A general formula, in case of non-magnetic scatterer ($\mu=1$), linking these two fields is~\cite{yurkin_review_2007}:

\begin{equation}
    \mathbf{E}_i^\ell=[\overline{\mathbf{I}}+(\overline{\mathbf{L}}_i-\overline{\mathbf{M}}_i)\overline{\boldsymbol{\chi}}_i]\mathbf{E}_i.
    \label{eq:Emicro_to_Emacro}
\end{equation}
Here $\overline{\mathbf{L}}_i$ is the self-term dyadic, which equals to $\overline{\mathbf{I}}/3$ for cubical (or spherical) voxels, and $\overline{\mathbf{M}}_i$ is the finite-size correction, which choice determines the specific polarizability prescription. For instance, assumption of $\overline{\mathbf{M}}_i=0$ corresponds to the so-called weak form of the DDA~\cite{lakhtakia_weak_1992}, leading to:
\begin{equation}
    \mathbf{E}_i^\ell = \mathbf{E}_i\frac{\overline{\boldsymbol{\varepsilon}}+2\overline{\mathbf{I}}}{3}.
\end{equation}
and thus to Clausius--Mossotti polarizability.

Note that all three codes return the total field as output, while \ifdda also provides the local field. However, the initialization of an iterative solver may differ. While \ifdda always defines the initial field in terms of the local field, \adda can define in terms of the local or total field depending on the \cmd{-init_field} mode (see Sec.~12.1 of its manual~\cite{ADDAmanual_2020}).
Moreover, \ifdda scales this field using the Hegedüs trick~\cite{Strakos_hagedus_2005}.

This leads in matrix notation to:
\begin{equation}
(\mathbf{I} - \mathbf{G} \mathbf{D}_\alpha) \, \mathbf{E}^\ell = \mathbf{E}^\mathrm{inc}.
\end{equation}
Here $\mathbf{D}_\alpha = \mathrm{diag}(\overline{\boldsymbol{\alpha}}_j)$ is a block-diagonal matrix of polarizabilities.
\medskip
To be consistent with the two previous forms, we can define a matrix $\mathbf{A}^E = \mathbf{I} - \mathbf{G} \mathbf{D}_\alpha$, where the elements of the matrix are:
\begin{equation}
\overline{\mathbf{A}}^E_{ij} = 
\begin{cases}
\mathbf{I}, & i = j \\
-k^2\overline{\mathbf{G}}_{ij} \, \overline{\boldsymbol{\alpha}}_j, & i \neq j
\end{cases}
\end{equation}
This matrix $\mathbf{A}^E$ always has unit diagonal but is complex symmetric only when all $\overline{\boldsymbol{\alpha}}_i$ are identical scalars. 

\subsubsection{Relationship between $\mathbf{A}^x$, $\mathbf{A}^E$, $\mathbf{A}^p$}
For a homogeneous and isotropic target, the polarizability reduces to a scalar $\overline{\boldsymbol{\alpha}}_i = \alpha \mathbf{I}$ for all voxels.
In this special case, the three matrices reduce to
\begin{equation}
\mathbf{A}^x = \mathbf{A}^E = \alpha\mathbf{A}^p,    
\end{equation}
which differ only by a constant scaling. However, the linear systems also differ by the choice of primary unknown
($\mathbf{E}^\ell$, $\mathbf{p} = \overline{\boldsymbol{\alpha}} \mathbf{E}^\ell$,
and $\mathbf{x} = \overline{\boldsymbol{\beta}}^{-1} \mathbf{E}^\ell$) and source terms ($\mathbf{E}^\mathrm{inc}$ and $\mathbf{b}=\overline{\boldsymbol{\beta}}\mathbf{E}^\mathrm{inc}$), with
$\overline{\boldsymbol{\beta}} = \overline{\boldsymbol{\alpha}}^{1/2}$. Thus, when evaluating the residual:
\begin{equation}
    \mathbf{r}=\frac{\|\mathbf{b}-\mathbf{A}\mathbf{x}\|}{\|\mathbf{b}\|}
\end{equation}
of the systems, the scalar factor cancels out, resulting in identical solutions up to machine precision, with only minimal differences caused by round-offs and the order of operations.

For anisotropic or inhomogeneous materials, however, the polarizability is no longer a scalar and may vary from voxel to voxel, which breaks commutativity of $\mathbf{G}$ and $\mathbf{D}_\alpha$ (or $\mathbf{D}_\beta$). As a result, $\mathbf{A}^x$, $\mathbf{A}^E$, and $\mathbf{A}^p$ are no longer related by simple scalings, and their residuals cannot be reconciled merely by transformations of the primary unknowns or source terms. Nevertheless, all systems aim to approximate the same physical scattering problem, and since the DDA is a deterministic method, only one exact solution exists. The differences between the computed solutions are therefore dominated by the solver tolerance, resulting in $\eta$-level agreement. 

\adda supports anisotropic materials but only under the assumption that the dielectric tensor of the target is diagonal in the chosen frame. 
\ddscat, on the other hand, can handle off-diagonal components by constructing a rotated dielectric tensor: the user provides the diagonal elements (in the local dielectric frame) along with rotation angles, and \ddscat computes the off-diagonal components (see Sec.~28 of the manual~\cite{draine2013userguidediscretedipole}). However, it assumes that the dielectric tensor is symmetric. 
In contrast, \ifdda allows for a fully general dielectric tensor, including off-diagonal elements, however, anisotropic materials are implemented only in combination with Clausius--Mossotti or radiative reaction correction~\cite{draine_discrete_1988} polarizabilities. Naturally, all three codes support inhomogeneous materials.

\subsection{Impact of free parameters on simulation agreement}
\label{sec:free_parameters}

Even though DDA is a deterministic method, small differences in free parameters can strongly affect the numerical agreement between independent runs. To quantify this sensitivity, we varied individual parameters between two otherwise identical DDA simulations and compared the resulting extinction efficiencies and internal fields. The number of matching digits between the two results serves as a practical metric of agreement, where 10--15 digits correspond to full machine precision for doubles. Table~\ref{tab:parameter_accuracy} summarizes the expected level of agreement for each parameter category when only that parameter differs across codes or configurations.

\begin{table}[!ht]
\centering
\footnotesize
\begin{threeparttable}
\caption{Expected agreement (in terms of matching digits) when varying individual parameters across DDA codes.}
\label{tab:parameter_accuracy}
\begin{tabularx}{\textwidth}{lXl}
\toprule
Parameter category & Description & Matching digits \\
\midrule
Studied system & particle shape, refractive index, wavelength & 0 \\
Grid & number and size of voxels & 1--2\tnote{a} \\
DDA formulation & polarizability model and interaction formulation (e.g., LDR, FCD, IGT) & 1--2 \\
Solver parameters & type of iterative solver and convergence threshold $\eta$ & $\eta$-level \\
Initial field & initial guess for the iterative solver (e.g., zero, WKB, or preconditioned) & $\eta$-level \\
\bottomrule
\end{tabularx}
\begin{tablenotes}
\footnotesize
\item[a] This depicts the typical value of method error (discussed in Sec.~\ref{subsec:accuracy_dda}), which can also be significantly smaller.
\end{tablenotes}
\end{threeparttable}
\end{table}

As shown in Table~\ref{tab:parameter_accuracy}, discrepancies in the considered scattering problem immediately break reproducibility, yielding, in general, zero matching digits. Differences in the DDA formulation or computational grid typically yield one or two significant digits, where the number of digits is quantified such that a relative error of $10^{-x}$ corresponds to $x$ digits of agreement. While solver-related parameters (choice of algorithm, stopping criterion $\eta$, or initial field) generally control the agreement level at the $\eta$-digit level. For example, with $\eta = 10^{-4}$, agreement is typically limited to four digits.

Full machine-precision agreement can be reached even at the $\eta$-level, provided that the solver tolerance is tightened to $10^{-16}$ in double precision.

\subsection{Default parameters and unified command-line setup}
\label{sec:default_parameters}

To enable floating-point--consistent cross-verification between independent DDA codes, all relevant numerical and physical parameters must be harmonized. Although \ifdda, \adda, and \ddscat share a similar theoretical foundation, their default configurations differ significantly in terms of target definition, polarization handling, numerical conventions, and solver parameters. Table~\ref{tab:default_parameters} summarizes the main default values for each code. These differences must be explicitly resolved to achieve consistent and reproducible simulations.

While all three implementations employ a plane-wave incident field by default, differences remain in polarization handling and coordinate conventions. Both \adda and \ddscat can compute two orthogonal polarizations simultaneously, whereas \ifdda currently performs one polarization per
run. The propagation and polarization directions also differ by default. Moreover, polarizability models and solver choices differ substantially, leading to variations in iterative convergence and accuracy if not explicitly aligned.

\begin{table}[!ht]
\centering
\tiny
\begin{threeparttable}
\caption{Default parameters across DDA codes. Differences must be resolved to achieve machine-precision equivalence.}
\label{tab:default_parameters}
\begin{tabularx}{\textwidth}{lXXl}
\toprule
Parameter & \ifdda & \adda & \ddscat \\
\midrule
Shape & sphere & sphere & 16-sphere aggregate \\
Size & $R=\qty{100}{nm}$ & $D\approx \qty{6.734552}{\um}$\,\tnote{a} & $a_\mathrm{eff}=\qty{0.25198}{\um}$ \\
Incident field & plane wave & plane wave & plane wave \\
Propagation direction & $\hb{z}$ & $\hb{z}$ & $\hb{x}$ \\
Polarization direction & $\hb{y}$ & $\hb{y}$ & $\hb{y}$ \\
Target orientation & $zxz$-convention & $zyz$-convention & $xzx$-convention \\
Wavelength $\lambda$ & \qty{632.8}{nm} & $2\pi$\,\unit{\um} & \qty{0.6}{\um} \\
Refractive index & $\varepsilon=1.1$ & $m=1.5$ & $m=1.33 + 0.01i$ \\
Polarizability & RR & LDR & CLDR \\
Grid resolution $n_x$ & 10 & 16 & 24 \\
Solver & GPBiCG1 & QMR & PBCGS2 \\
Convergence threshold $\eta$ & $10^{-4}$ & $10^{-5}$ & $10^{-5}$ \\
Initial field & Born & automatic & zero field \\
Volume correction & none & enabled & enabled \\
Scattering formula & Draine's formulation~\cite{draine_discrete_1988} & Draine's formulation & Draine's formulation \\
Unit system & SI outputs (with code equations in CGS) & CGS & CGS \\
\bottomrule
\end{tabularx}
\begin{tablenotes}
\footnotesize
\item[a] Equal to $(6 N/\pi)^{1/3}\lambda/(10|m|)$ for default values of $\lambda$ and $m$, while $N=2176$ for a sphere with the default value of $n_x$.
\end{tablenotes}
\end{threeparttable}
\end{table}

To facilitate systematic benchmarking, we developed a Python wrapper enabling \ddscat to be executed directly from the command line~\cite{Argentin_ddscatcli_2025}.
Using it we can formulate minimal command-line examples constructed with consistent physical and numerical parameters, yielding agreement up to machine precision between the three codes:
\begin{lstlisting}[style=cli]
$./adda -lambda 632.8 -size 200 -init_field zero -grid 15
$./ifdda -polarizability LS -methodeit QMRCLA -epsmulti 2.25 0.0 -nnnr 15 -ninitest 0 -tolinit 1.d-5 -object sphere 99.5540512465
$ddscatcli -CMDSOL QMRCCG -CALPHA LATTDR -MEM_ALLOW "15 15 15" -CSHAPE ELLIPSOID -SHPAR "15 15 15" -DIEL "diel/m1.50_0.00" -WAVELENGTHS "632.8 632.8 1 'LIN'" -AEFF "100 100 1 'LIN'" -run
\end{lstlisting}
Note that it is still up to the user to manually create or configure the input file for the refractive index in \ddscat. In the above and following command lines, we use the \ddscat style convention to name them according to their value.

A table with all the equivalent parameters and their corresponding accuracy agreement is provided in \ref{Appendix:equivalence}. For all command-line examples, we exploit the fact that \ddscat and \adda allow for arbitrary physical units for all lengths (consistently for all input and output) to express the input wavelength and particle size in nanometers to match the convention, postulated by \ifdda. This choice is made solely to improve readability and usability and does not affect the iterative convergence process. Moreover, the equivalence between the codes and the scale-invariance rule \cite{MISHCHENKO2006scale} in general indicates that the same flexibility exists for \ifdda, although some care should be exercised in interpreting the output. For instance, the resulting cross sections would not have the units of \unit{m^2} if the input unit is not \unit{nm}.

\section{Performance comparison: CPU benchmark}
\label{sec:comparison}

This section presents the performance benchmarks conducted across the three DDA implementations: \ifdda, \adda, and \ddscat. CPU benchmarks are performed for all three codes, while GPU benchmarks are restricted to \adda and \ifdda. The main goal is to evaluate runtime, memory footprint, and precision effect under identical physical and numerical conditions. Notably, \ddscat and \ifdda allow single-precision execution, respectively on CPU and GPU, providing additional insight into performance–accuracy trade-offs.

To ensure fair comparisons, all benchmarks were configured to achieve machine-precision agreement across the three implementations. This requirement imposes tight constraints on the simulation setup, as not all combinations of parameters or solver options are available in every code. We therefore focus on a single, well defined case, a homogeneous cubic particle of ice with $kD = 30$ ($D$ is the edge size), solved using the BiCGStab iterative solver and the FCD interaction and polarizability formulation. This configuration is representative of typical DDA workloads and was also used in Ref.~\cite{penttila_comparison_2007}.

As mentioned in Sec.~\ref{sec:overview}, BiCGStab is chosen to avoid the two times difference in algorithmic complexity between QMR implementations in different codes. However, BiCGStab implementations also slightly differ in the stopping procedure. The original algorithm~\cite{BiCGStab_original} can stop in the middle of iteration, saving one matrix-vector product. This is implemented in \adda, but not in the PIMZ~\cite{cunha_parallel_1995} version of \ddscat and \ifdda. Whether this happens is quasi-randoms depending on problem parameters. The following test case was explicitly tuned to avoid this issue. Other scattering problems may, instead, lead only to $\eta$-level of agreement and extra computational-time difference for a single matrix-vector product.
Hence, unfortunately, there are currently no ideal iterative solver for robust comparison between all three codes. However, such options exist for pair-wise comparisons (see Table~\ref{tab:cli_equivalences_levels}).

Due to floating-point rounding errors, the convergence history slightly differ between the codes and hardware backends, leading to different iteration counts for a fixed~$\eta$. This degrades floating-point consistency at the $\eta$-level.
To remove this issue, we explicitly tuned~$\eta$ for each implementation so that all CPU runs terminated after the same number of iterations (54 in this case), yielding 10--12 matching digits in the extinction cross section and efficiency (see Table~\ref{tab:matching_digits}). 

An alternative and more robust approach to enforce identical iteration counts would be to limit the maximum number of iterations using options such as \texttt{-maxiter} (in \adda), \texttt{-MXITER} (in \ddscat), or \texttt{-nlim} (in \ifdda). However, while \adda still computes physical observables when this limit is reached, both \ifdda and \ddscat terminate execution, preventing a comparison. We therefore adopt threshold tuning as a practical compromise, however, we suspect such tuning to become impractical beyond 100 iterations.

Since our comparison focuses on runtime and solver performance rather than angular quantities, we reduced the angular sampling: in \ddscat we set \texttt{ETASCA = 10} as recommended by the manual and used \texttt{IORTH = 1} to compute a single polarization; for \adda this corresponds to \texttt{-ntheta 10}. The exact command lines used were:

\begin{lstlisting}[style=cli]
$./ifdda -object cube 2387.3241463784303 -lambda 500 -epsmulti 1.723969 0.0 -ninitest 0 -nnnr 150 -tolinit 9.46237161365793d-5 -methodeit BICGSTAB -polarizability FG
$./adda -shape box 1 1 -size 2387.3241463784303 -lambda 500 -m 1.313 0.0 -init_field zero -grid 150 -eps 4.024 -iter bicgstab -pol fcd -int fcd -scat dr -ntheta 10 -maxiter 54
$ddscatcli -CSHAPE RCTGLPRSM -AEFF "1480.9777061418503 1480.9777061418503 1 'LIN'" -WAVELENGTHS "500 500 1 'LIN'" -DIEL "diel/m1.313_0.00" -MEM_ALLOW "150 150 150" -SHPAR "150 150 150" -TOL 9.46237161365793e-5 -CMDSOL PBCGST -CALPHA FLTRCD -ETASCA 10 -IORTH 1 -NPLANES 0 -NRFLD 0 -run
\end{lstlisting}

The memory footprint grows linearly with the number of voxels $N$, so it can be extrapolated straightforwardly for each code. By contrast, the reported runtimes and speedups in this chapter are problem-size dependent. As discussed in Refs.~\cite{donald2009opendda,yurkin_review_2023}, the time per iteration does not scale exactly linearly with $N$ (e.g., FFT introduce $N\log N$ behavior) and depends sensitively on the FFT implementation, numerical precision, and the paradigm of parallelization. These effects are also hardware-dependent.

A comprehensive sweep over \(N\) is beyond our present study. Consequently, the speedups reported here characterize the selected test cases and hardware, and should be interpreted or extrapolated with caution to different problem sizes.

\subsection{Computational Setup}
The CPU architectures used for benchmarking are summarized in Table~\ref{tab:hw-summary}. They correspond respectively to a single node of a high-performance Cray cluster and to a modern laptop processor. This contrast allows us to illustrate both the performance potential and the practical limitations that can be expected on HPC hardware versus everyday desktop environments.

\begin{table}[!ht]
\centering
\caption{CPU platforms used in benchmarks. Cache sizes are aggregated \emph{per socket}. Memory bandwidths are theoretical peaks \emph{per socket} (decimal \unit{GB/s}).}
\label{tab:hw-summary}
\footnotesize
\begin{tabular}{@{}lcc@{}}
\toprule
 & AMD EPYC 9654 (Genoa) & Intel Core Ultra 7 165H \\
\midrule
Sockets & 2 & 1 \\
Cores / Threads & 96 / 192 (per socket) & 16 / 22 (8 E + 6 P cores) \\
Base / Turbo (GHz) & 2.4 / 3.7 & 1.4 / 5 \\
L1I / L1D & \qty{3}{MiB} / \qty{3}{MiB} & \qty{896}{KiB} / \qty{544}{KiB} \\
L2 / L3 & \qty{96}{MiB} / \qty{384}{MiB} & \qty{18}{MiB} / \qty{24}{MiB} \\
Peak memory BW & \qty{460.8}{GB/s} (DDR5-4800) & \qty{120}{GB/s} (LPDDR5x-7467) \\
RAM (node) & \qty{768}{GiB} ($24 \times \qty{32}{GiB}$) & $2 \times \qty{32}{GiB}$ \\
\bottomrule
\end{tabular}
\end{table}

All cluster executions were submitted using the \texttt{SBATCH --exclusive} option of SLURM to eliminate performance variability due to shared-node contention. Each simulation was repeated three times to compute the mean and sample standard deviation, assessing run-to-run variability. 

All three codes were compiled using the same toolchain to minimize environment-related bias. Compiler version had negligible impact: switching from \texttt{GCC}~11.2 to \texttt{GCC}~12.2 produced identical results within measurement noise. On the cluster, we used the following software stack: \texttt{GCC}~12.2 with \texttt{OpenMP}~4.5, \texttt{MPICH}~3.4a2, \texttt{FFTW}~3.3.10, and \texttt{Intel oneMKL}~2023.0, while on the laptop: \texttt{GCC}~11.4 with \texttt{OpenMP}~4.5, \texttt{Open MPI}~4.1.2, \texttt{FFTW}~3.3.8, and \texttt{Intel oneMKL}~2025.0. \adda was compiled with \texttt{-Ofast} optimization flag, \ifdda with \texttt{-O3}, and \ddscat with \texttt{-O2}, following their respective build defaults.

The following subsections present the CPU results while the GPU benchmark is reported separately in another section. Each test reports total runtime, speedup factors, memory footprints, and precision effects.

\subsection{Time Performance}
\label{sec:cpu_times}

We tested several grids and FFT factorizations and found that grid sizes of $n_x = 150$ and $n_x = 250$ allow all codes to show their best. For all CPU benchmarks, the number of MPI processes was chosen to follow \adda recommended practice for parallel FFTs, i.e., it divides $n_x$. Figures~\ref{fig:fig1_dda_runtime_speedup} and~\ref{fig:fig2_dda_runtime_speedup} report wall-clock times and speedups on the AMD EPYC\,9654 node and the Intel Core Ultra 7\,165H processor, respectively. Error bars denote the sample standard deviation and are shown only when the corresponding coefficient of variation exceeds 5\%. 

In this section, we report only the total wall-clock time, since it is dominated by the iterative solver phase, which is itself dominated by matrix-vector products, i.e., FFTs. By contrast, for GPU benchmarks, the solver phase is the portion offloaded to the device and, thus, its time may become comparable to that of the parts remaining on the CPU.

\begin{figure}[!ht]
    \centering
    \includegraphics[width=\linewidth]{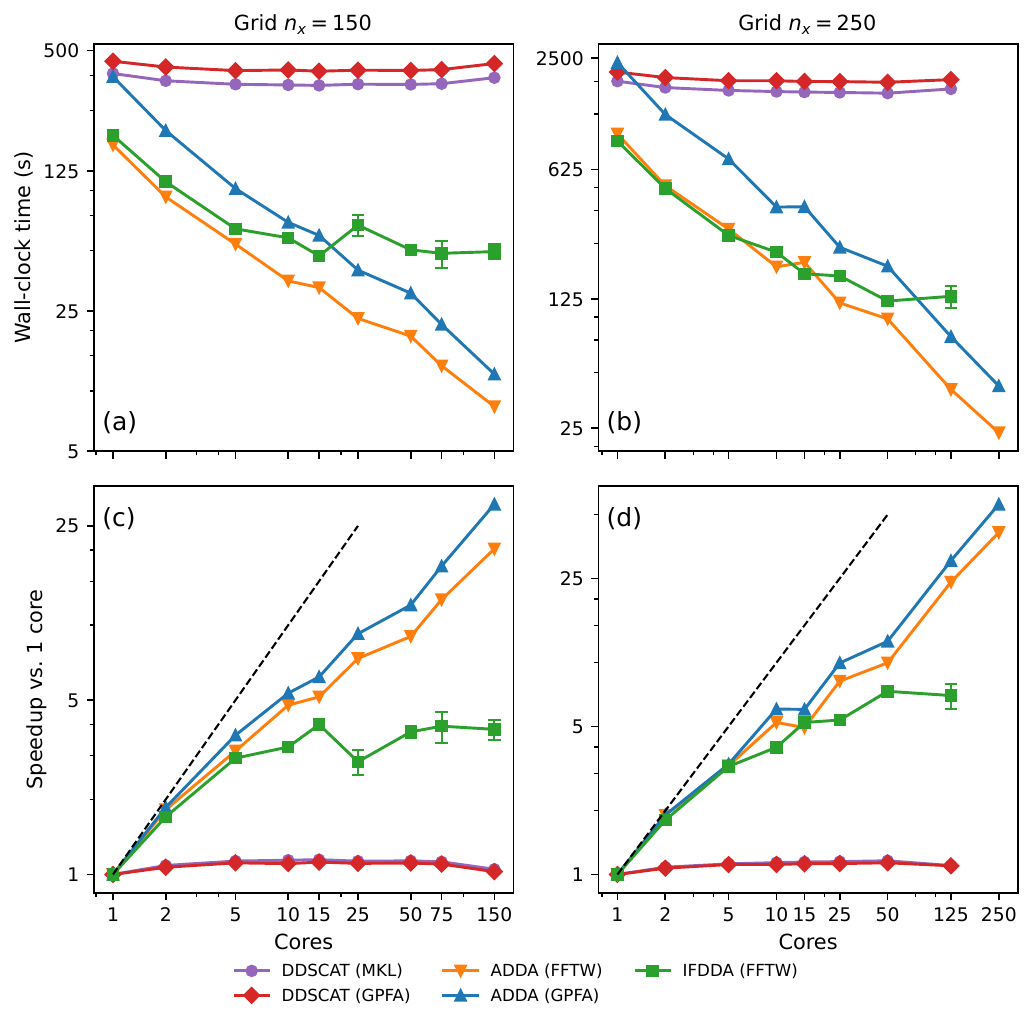}
    \caption{
    Wall-clock time (top row) and speedup relative to the single-core reference (bottom row), for grid sizes $n_x=150$ (left) and $n_x=250$ (right) on the AMD EPYC\,9654 node. Log-log scale is used, and the dashed curves indicate ideal linear scaling. Error bars, denoting the sample standard deviation, are shown when it is larger than 5\% of the value.
    }
    \label{fig:fig1_dda_runtime_speedup}
\end{figure}

Consistent with the \ddscat manual~\cite{draine2013userguidediscretedipole}, MKL outperforms the legacy GPFA backend on modern CPUs, especially on Intel processors, as MKL is heavily optimized for Intel microarchitectures. As a result, \texttt{DDSCAT MKL} achieves roughly twice the performance of its GPFA build on the Intel laptop, while the improvement is more modest on the AMD EPYC node. This makes \texttt{DDSCAT MKL} competitive with \texttt{ADDA FFTW} and \texttt{IFDDA FFTW} in single-core runs on the Intel laptop for the larger grid. On the AMD node, however, \texttt{ADDA FFTW} and \texttt{IFDDA FFTW} remain significantly faster, outperforming \texttt{DDSCAT MKL} by factors of roughly $2.5$ and $2.0$ for $n_x = 150$ and $n_x = 250$, respectively. These differences are expected. Indeed, while \ddscat performs a full 3D FFT transform in one run~\cite{Goodman1991FFT}, \adda reduces the total FFT workload by decomposing the 3D convolution into a sequence of 1D FFTs with intermediate transpositions. This 3D-to-1D strategy, introduced in Ref.~\cite{Hoekstra19981dfft}, almost halves the number of involved 1D FFTs by explicit usage of zero-padding inherent to the convolution, sustaining high parallel efficiency and optimizing both memory usage and execution time. During the course of this work, \ifdda was also modified to adopt a similar 3D-to-1D implementation as \adda, but optimized primarily for execution time (see Sec.~\ref{sec:memory} for further details on memory differences), resulting in performance gains of approximately 10\% here.
Related improvements also appear in \texttt{OpenDDA}~\cite{donald2009opendda}. Finally, the FFTW library, featuring hardware-adaptive planning, shows optimal performance over all tested hardware, in agreement with previous inter-code comparisons~\cite{penttila_comparison_2007}. 

For multi-core CPU runs, \ddscat exhibits essentially flat scaling, as its FFT routines, which constitute the largest fraction of the total runtime, are not parallelized with OpenMP. \ddscat parallelization is primarily designed for orientation-averaging (outside of our scope), where much larger speedups would be observed. By contrast, both \adda and \ifdda scale reasonably well on the AMD node, achieving speedups up to 24$\times$ (for 125 cores) and 7$\times$ for $n_x = 250$. On this architecture, \texttt{ADDA FFTW} remains more than $5\times$ faster than \texttt{IFDDA FFTW} for $n_x=150$ and more than 3$\times$ faster for $n_x=250$ (for 125 cores). Moreover, we observed that variability is greater for OpenMP than for MPI. On the Intel laptop, however, \ifdda displays better runtimes and scaling for $n_x=250$ reaching speedup up to 1.5$\times$.

\begin{figure}[ht!]
    \centering
    \includegraphics[width=\linewidth]{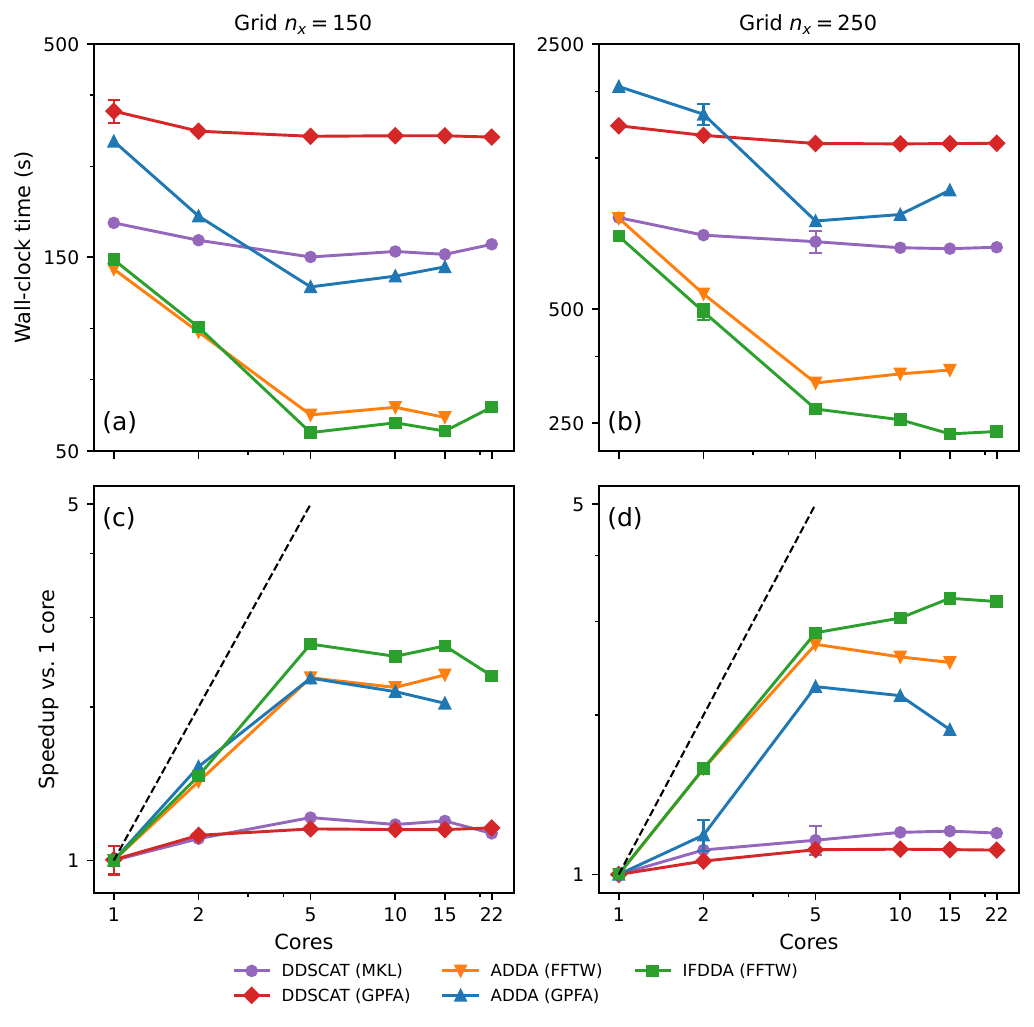}
    \caption{
    Wall-clock time (top row) and speedup relative to the single-core reference (bottom row), for grid sizes $n_x=150$ (left) and $n_x=250$ (right) on the Intel Core Ultra 7 165H processor. Log-log scale is used, and the dashed curves indicate ideal linear scaling. Error bars, denoting the sample standard deviation, are shown when it is larger than 5\% of the value.
    }
    \label{fig:fig2_dda_runtime_speedup}
\end{figure}

To explain these behaviors it is necessary to discuss saturation effects. As visible in Fig.~\ref{fig:fig1_dda_runtime_speedup}, \adda and \ifdda exhibits nearly linear scaling up to 2--3 cores, after which deviations from ideal scaling appear. This saturation arises from Amdahl's law and memory-bandwidth limits, as more threads compete for a fixed number of memory channels, communication costs progressively decrease parallel efficiency. Similar trends were reported in a recent comparative study of FFTW backends using OpenMP and MPI~\cite{strack2025parallelfftwriscvcomparative}. The different saturation points between $n_x = 150$ and $n_x = 250$ highlight the grid-dependent scaling behavior of 3D FFTs. Larger grids maintain high parallel efficiency for longer, as their computational intensity grows faster than synchronization and communication overheads.

On the Intel laptop, these effects appear much sooner. The available memory bandwidth is significantly lower and the cache hierarchy is smaller (see Table~\ref{tab:hw-summary}), causing the working set of the 3D FFT to overflow to main memory. As a result, speedups saturate well before 10 cores despite slightly better single-core performance due to higher clock frequency. This also explains why \texttt{ADDA GPFA} exhibits better scaling than \texttt{ADDA FFTW} on the AMD node, while the opposite trend is observed on the Intel laptop (although the wall time of FFTW is superior in all cases). Specifically, the limited memory bandwidth and cache capacity make the adaptive nature of FFTW more beneficial during the parallel execution on the Intel laptop. This highlights the strong dependence of FFT performance on both the underlying hardware architecture and the algorithmic optimization strategy.

Finally, note that the wall-clock time of \texttt{ADDA FFTW} for $n_x = 250$ at 250 cores was obtained using two AMD nodes, as a single EPYC\,9654 node provides only 192 cores. This was done to illustrate that \adda MPI implementation continues to scale across nodes. The runtime can decrease further when using more nodes with fewer cores per node, indicating that latency effects remain a limiting factor. A more systematic study, including a saturation predictor based on grid size and hardware characteristics, would be valuable for future work.

To conclude, users should prefer FFTW or MKL over GPFA backends and, for large-scale problems, use the MPI execution of \adda.

\subsection{Precision effects}
\label{sec:cpu_single}

A practical advantage of \ddscat is that it supports single precision, which reduces the array footprints by a factor of two. For $n_x=150$ case, single precision also reduces the per-iteration time since the iteration counts matched the double precision one.

Switching from double to single precision yielded a time reduction of approximately 30\% with GPFA and 60\% with MKL on the laptop, and 10\% and 40\% respectively on the cluster (data not shown). Moreover, accuracy remained good, as results match the $\eta$-level digits.

Single precision can therefore be faster than \ifdda and \adda on a single core of the laptop for this grid (115 seconds with MKL). Overall, single precision in \ddscat offers substantial savings in memory (factor of two) and runtime (up to 60\% with MKL) with negligible effect on accuracy for this case. However, the latter is not guaranteed, especially with increasing number of iterations. Thus, more sophisticated approaches to use the double precision for only part of variables have been proposed~\cite{penttila_comparison_2007}. Users should therefore be cautious when trading performance gains against potentially reduced accuracy.

\subsection{Memory performance}
\label{sec:memory}

Table~\ref{tab:memory-grid150} reports memory footprints using binary prefixes as recommended by the International Electrotechnical Commission (IEC): \unit{GiB} denotes gibibytes ($\qty{1}{GiB}=2^{30}\,\unit{B}$). The only exception is memory bandwidth in Table~\ref{tab:hw-summary}, which uses decimal \unit{GB}. We report physical memory only, i.e., the resident set size (RSS). In our experience the virtual memory size (VM/VSZ) is not a useful capacity indicator for these workloads: VM is always $\ge$\,RSS because it includes the full address space (reserved but untouched pages) and the complete size of shared libraries mapped into the process; by contrast, RSS counts only the pages that are actually resident in RAM.

\begin{table}[!ht]
\centering
\caption{Node-level physical memory [\unit{GiB}] for grid size $n_x=150$.
Totals are RSS per node; for MPI (\adda), computed as the average of RSS memory from each rank, times the number of ranks (from \texttt{sacct}).}
\label{tab:memory-grid150}
\begin{tabular}{l*{5}{c}}
\toprule
\textbf{Method} & \textbf{1} & \textbf{25} & \textbf{50} & \textbf{75} & \textbf{150} \\
\midrule
\adda        & 2.5 & 3.0 & 4.2 & 5.4 & 12.5 \\
\ifdda       & 6.5 & 6.5 & 6.5 & 6.5 & 6.6 \\
\ddscat      & 6.3 & 6.3 & 6.3 & 6.3 & 6.3 \\
\bottomrule
\end{tabular}
\end{table}

For codes that use OpenMP (\ifdda and \ddscat), threads share a single process and therefore one copy of the large working arrays; RSS remains essentially flat as the thread count increases (see Table~\ref{tab:memory-grid150}). For an MPI program, however, an increase is observed with the number of ranks: although \adda uses about 2.5 less memory at one rank, the aggregate memory can exceed the OpenMP codes at higher core counts. This behavior is primarily due to MPI. Compared to shared memory, MPI launches $P$ separate processes. Each rank keeps its own copy of certain auxiliary MPI variables, leading to a noticeable memory increase with the number of ranks. This MPI-related overhead is clearly visible in moderate-size runs but becomes negligible for larger grids, where the properly distributed working arrays dominate. A detailed analysis of the MPI memory scaling is provided in~\ref{Appendix:mpi_memory}.

The lower initial memory usage in \adda originates from how the interaction matrix is stored in Fourier space. Indeed, in FFT-accelerated DDA formulations, only the six independent components of the dyadic Green's tensor are required due to symmetry; these components are transformed to Fourier space and stored on an extended zero-padded domain of twice the size in each spatial direction. In the classical approach~\cite{Goodman1991FFT}, all six components are stored over the full circulant domain. \adda reduces this cost by exploiting mirror symmetries in two directions, while not using symmetry along the third axis to facilitate MPI parallelization, thereby reducing the storage requirement by a factor of four compared to the classical formulation. Further details are provided in Ref.~\cite{donald2009opendda}, which also shows how OpenDDA exploits symmetry in all three spatial directions to reach the theoretical minimum storage.

On the full node (192 cores, \qty{714}{GiB} RAM) we were able to run ADDA up to $n_x=960$. Moreover, using fewer cores per node and moving to multiple nodes allow even larger grids. For example, a recent large-scale ADDA run was reported in Ref.~\cite{YurkinELSC2025}, reaching $n_x=3168$ for a cube with size parameter $kD=2000$ and refractive index $m=1.02$, using 44 nodes (3168 CPU cores) with a peak memory footprint of \qty{27.6}{TiB} and an MPI overhead of only \qty{0.4}{TiB} (i.e., 1.2\%).

\section{Performance Comparison: GPU Benchmark}
\label{sec:comparison_gpu}

The GPU modes of both \ifdda (v1.0.27) and \adda (v.1.5.0-alpha3, commit \texttt{b03d648}) are still under active development. Beyond version control, the ongoing development activity is also evident from the public \adda issue tracker and recent benchmark discussions on the official discussion group. It is also worth noting the recent fork \texttt{ADDA-CXX}\footnote{\url{https://github.com/michelgross34/adda-CXX}} by one of the authors, which converts the original C99 codebase to C++ to facilitate future CUDA integration (not used in this work). The timings and memory footprints reported here therefore characterize the current implementations and test configurations rather than hard limits. Thus, future releases will significantly change these results. 

\subsection{Computational Setup}
The GPU benchmarks were performed on four NVIDIA accelerator devices spanning both cluster- and workstation-architectures. These include the A100 (Ampere), H200 (Hopper), RTX~6000 Ada, and RTX~2000 Ada. Their main architectural characteristics are summarized in Table~\ref{tab:gpu-summary}.

Recent NVIDIA GPUs integrate tensor cores units specialized in matrix–multiply–accumulate (MMA) that accelerate general matrix multiplication (GEMM) operations. They deliver very high throughput in mixed/low precision but also in standard precision (FP64 tensor core) on recent A100/H100 GPUs. However, standard FFT libraries (FFTW, cuFFT, clFFT) do not use them by default. Research prototypes that reformulate FFTs to be tensor cores-friendly~\cite{li2021tcfft} reported 1.1--3$\times$ speedups over cuFFT on NVIDIA A100, but we have not considered them.

\begin{table}[!ht]
\centering
\caption{GPU platforms used in the benchmarks. Peak floating-point throughput and memory bandwidth values are vendor-reported, while cache sizes are taken from third-party hardware databases. FP64 throughput on Ada GPUs is $1/64$ of FP32 peak. SM denotes ``Streaming Multiprocessors''.}
\label{tab:gpu-summary}
\tiny
\begin{tabular}{@{}lcccc@{}}
\toprule
 & A100 (\qty{80}{GiB} SXM4) & H200 (\qty{141}{GiB} SXM) & RTX~6000 Ada & RTX~2000 Ada \\
\midrule
FP64 (\unit{TFLOP/s})         & 9.7       & 34     & 1.4 & 0.2 \\
FP32 (\unit{TFLOP/s})         & 19.5      & 67      & 91.1       & 12 \\
Mem.\ BW (\unit{GB/s}) & 2{,}039   & 4{,}890    & 960        & 256 \\
L1 / SM (\unit{KiB})   & 192       & 256       & 128        & 128 \\
L2 (\unit{MiB})        & 40        & 50        & 96         & 12 \\
SM                     & 108       & 132       & 142        & 22 \\
\bottomrule
\end{tabular}
\end{table}

Each GPU was paired with the host processor available on the corresponding system: dual-socket AMD EPYC 7543 nodes for the A100, dual-socket Intel Xeon 8558 nodes for the H200, an Intel Core Ultra 7 165H laptop for the RTX 2000 Ada, and an Intel Xeon w5--3525 workstation for the RTX 6000 Ada. Ideally, all GPUs would be benchmarked under the same host CPU to isolate accelerator performance. However, this was not possible due to hardware availability. Nevertheless, this does not affect the validity of our comparison, as for each run we separately record both the GPU-resident solver time and the total wall-clock time. This allows us to separate the contribution of the GPU from that of the host processor and to assess the impact of both GPU and CPU hardware on the overall runtime.

The same numerical configurations as in the CPU section were used to ensure direct cross-platform comparability. While the GPU double-precision runs also terminated in 54 iterations, the GPU single-precision runs with \ifdda converged in 53 iterations instead of 54. This difference arises because the step between successive residuals becomes too large to be tuned in the same manner as in double precision. Nevertheless, this does not affect the present comparison, since single-precision accuracy is inherently limited to 7--8 significant digits, and a difference of one iteration has a negligible impact on the measured execution times reported in Sec.~\ref{sec:comparison_gpu}. 

As in the CPU section, the reported timings correspond to the mean of three identical runs. Furthermore, all codes were compiled using the same library versions and optimization flags, with \texttt{CUDA}~13.0 for \ifdda and \texttt{OpenCL}~3.0 for \adda.

\subsection{Time performance}

For DDA solvers, the dominant cost per iteration is the FFT stage. Despite many optimization efforts over the years~\cite{barrowes2001fast,donald2009opendda,Shabaninezhad2021MPDDA}, FFTs remain the bottleneck of the iterative solver, with the remainder of the time spent mostly in BLAS (Basic Linear Algebra Subprograms) operations. In \ifdda, the entire solver loop is offloaded to the GPU, which minimizes host--device communication. In contrast, \adda offloads only the FFTs (matrix-vector products), while the linear-algebra operations remain on the CPU, introducing additional transfers. \adda also provides an experimental \cmd{OCL_BLAS} mode, which offloads the BLAS operations to the GPU via the \texttt{clBLAS} library, thereby both accelerating the BLAS operations and removing the corresponding transfers. As of the current \adda release, this mode supports only the BiCG solver. Since our focus in this study is on cross-code comparisons and floating-point--consistent reproducibility, we did not include it in the main benchmark. However, we report its performance in~\ref{Appendix:GPU}, where it achieves multiple speedups, demonstrating the benefit of a fully GPU-resident implementation.

We report in Fig.~\ref{fig:fig3_dda_runtime_gpu} solver-only time (iterative phase) and total wall-time (end-to-end) for grid sizes $n_x=150$ and $n_x=250$ on four NVIDIA GPUs (see Table~\ref{tab:gpu-summary}), except on the NVIDIA RTX~2000~Ada, where the IFDDA computation fails due to an out-of-memory error, and two CPU settings (1 and 10 cores). \ifdda is run in both double (DP) and single precision (SP), while \adda is reported only for 1 core because its default GPU mode is currently mutually exclusive with the MPI. For \adda, we also show the time for the the FFTs, as a part of the solver. In contrast to Fig.~\ref{fig:fig2_dda_runtime_speedup}, the sample standard deviation is not shown here for clarity. The corresponding coefficients of variation remain below 5\% in all cases (and typically around 1\%), except for the solver mean runtime of the \ifdda single-precision run for $n_x=150$ on the NVIDIA RTX 6000 Ada, where it reach 6\%.

\begin{figure}[!ht]
    \centering
    \includegraphics[width=\linewidth]{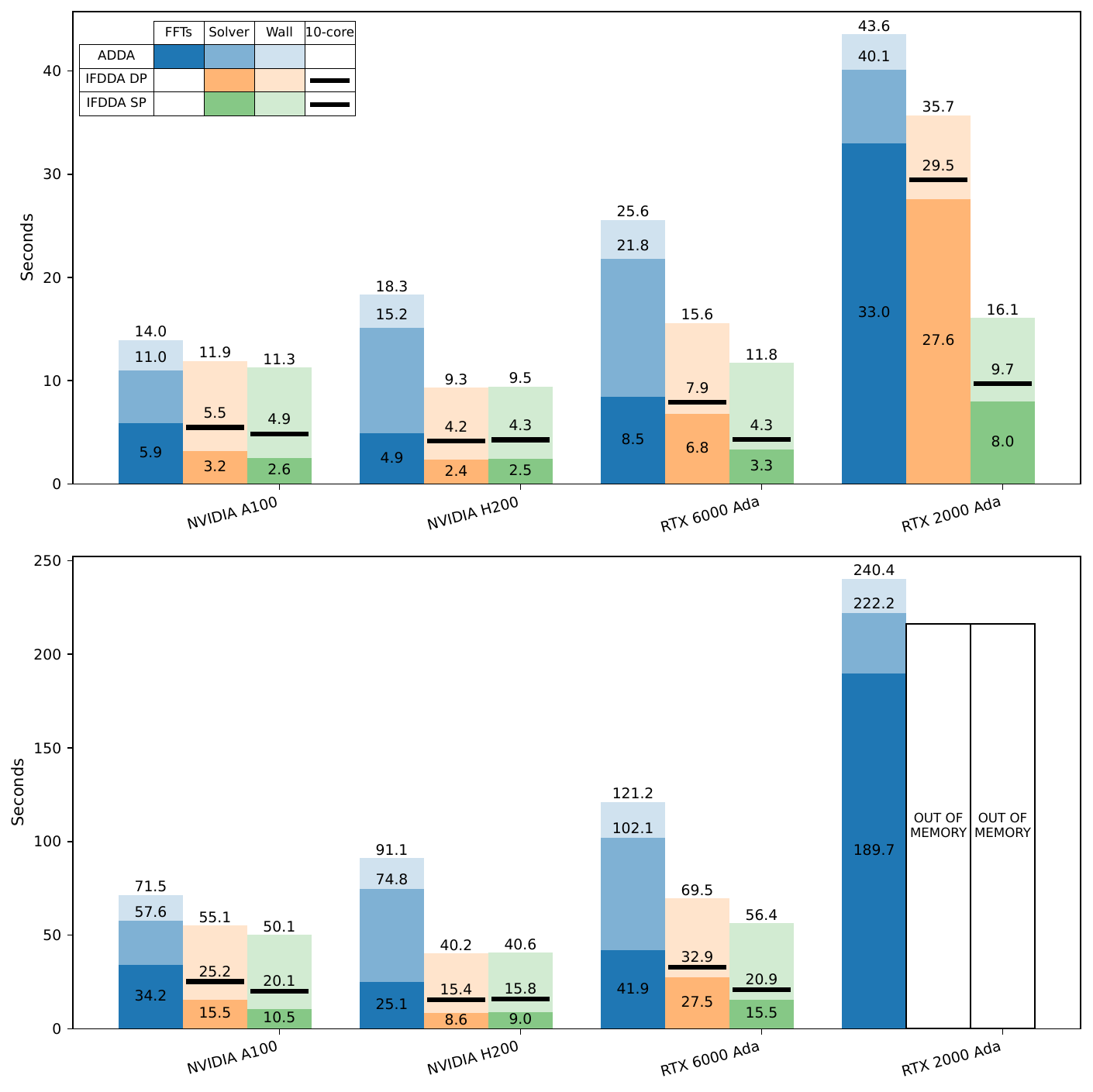}
    \caption{
    GPU timings for DDA codes. Bars are grouped by code and stacked to show FFT time, solver time, and $1$-core wall-time. Horizontal black lines indicate $10$-core wall-times for IFDDA (DP/SP). Grid size $n_x=150$ (top row); $n_x=250$ (bottom row).
    }
    \label{fig:fig3_dda_runtime_gpu}
\end{figure}

The overall GPU behavior of \texttt{IFDDA DP} can be interpreted within the roofline model~\cite{williams2009roofline}, which relates arithmetic intensity (\unit{FLOP/B}) to the machine balance (peak \unit{FLOP/s} per peak BW). The arithmetic intensity of 3D FFTs is low and depends on problem size; accordingly, DDA iterations lie predominantly in the memory-bound regime~\cite{ayala2022performance}, with possible transitions into the compute-bound or latency-bound regimes~\cite{williams2009roofline}. For a fully device-resident solver such as \ifdda, one therefore expects the solver time to scale approximately with the inverse memory bandwidth. The measured runtimes follow this trend qualitatively, as GPUs with higher bandwidth deliver shorter solver times, H200 $>$ A100 $>$ RTX~6000 Ada $>$ RTX~2000 Ada. However, quantitative description does not seem feasible, as synchronization overheads, cache and memory-path, and FP64/FP32 ratio vary significantly between devices. These factors introduce an additive latency term, causing a loss of performance. Consequently, the H200 does not reach the full improvement predicted by its bandwidth advantage over the A100 or workstation GPUs.

For \adda, the trend resembles that of \texttt{IFDDA DP} except for the H200, where the solver and total runtimes are higher than on the A100. Since only the FFTs are executed on the GPU, each solver iteration requires multiple host--device transfers and synchronizations, making the performance sensitive to system-level latency. Evidence for this interpretation is provided by the \cmd{OCL_BLAS} mode (see Fig.~\ref{fig:Appendix_fig1_ocl_blas}), where offloading the full solver to the GPU reverses the trend back to the expected one. This indicates that the performance inversion observed for \adda arises from latency costs in the hybrid CPU--GPU workflow, rather than from GPU-side throughput. We emphasize that it cannot be explained by any hardware specification, as both the memory hierarchy and CPUs associated with the H200 system are nominally superior (see Table~\ref{tab:gpu-summary}). Systematic study of this fact is left for the future.

In the single-core case, \texttt{IFDDA DP} systematically outperforms \adda, even when comparing only the GPU-offloaded part, which would in principle be expected to favor \adda since only FFTs are performed. This behavior can arise from several factors. First, the previously discussed CPU--GPU latency and synchronization overheads. Second, on NVIDIA hardware, \texttt{cuFFT} (used in \ifdda) is expected to outperform \texttt{clFFT}~\cite{steinbach2017gearshifft}. Finally, \adda relies on FFT plan transpositions using empirically chosen block sizes, which may be suboptimal for current GPU architectures. Taken together, these effects make it difficult to draw definitive conclusions or ratios from absolute runtimes alone. Similar considerations apply to total wall-clock times, where differences in CPU performance also contribute.

Nevertheless, by considering the relative speedup achieved by the \cmd{OCL_BLAS} mode on the solver part for the BiCG solver (see~\ref{Appendix:GPU}) and extrapolating this improvement to the BiCGStab case, one may expect \adda to achieve performance comparable to that of \ifdda when the entire solver loop is offloaded to the GPU. This interpretation is further supported by independent implementations. A MATLAB implementation, \texttt{MPDDA}~\cite{Shabaninezhad2021MPDDA}, which offloads the complete iterative solver to the GPU, reports better runtimes than the default \adda OpenCL mode, although this difference may partly arise from the use of Apple's OpenCL FFT backend rather than the standard \texttt{clFFT} library. More recently, a Python implementation, \texttt{CPDDA}~\cite{Xu2025CPDDA}, based on a similar strategy and relying on CuPy, reports performance surpassing \texttt{MPDDA} at smaller lattice spacings. Altogether, these results support the benefit of offloading the complete iterative solver to the GPU.

Finally, \ifdda can further reduce wall-clock time by increasing the number of CPU threads, even in GPU mode, since its non-FFT components benefit from OpenMP parallelism. For instance, using 10 CPU cores yields a $2\times$ reduction in the total runtime on the A100. 

\subsection{Precision Effects}
\label{sec:precision_effects}

For bandwidth-bound problems such as FFTs, switching from double to single precision ideally halves the memory traffic (8 to 4~bytes), potentially yielding a 2$\times$ speedup. For compute-bound regimes, the achievable ratio is instead limited by the device FP64:FP32 throughput, and for latency-bound regimes essentially no speedup is expected, since the dominant cost lies in waiting for data rather than processing it.

The measured ratios therefore differ across GPUs depending on their architectural balance between bandwidth, FP64 capability, and overheads. On the A100, FP64 performance is relatively strong (and half of FP32) and memory bandwidth is high. For moderate grid sizes ($n_x=150$--$250$) the kernels do not fully saturate the memory subsystem, and the observed speedups of 1.4--1.5$\times$ reflect a mixture of bandwidth and latency effects. For larger grids (not shown here), we observed the ratio of 1.8.

On the RTX~6000 Ada, the FP64 throughput is extremely low (of 1/64 FP32), and memory bandwidth is modest. This makes the latency delays less important, which explains larger DP/SP ratio of 1.8--2.1 still in the bandwidth-regime but closer to a compute-bound regime for $n_x=150$. This trend is even more pronounced on the RTX~2000 Ada, whose FP64 capability is still lower. There, the DP solver is clearly more compute-bound, manifested by the DP/SP ratio of 3.5, between bandwidth- and computations-based values.

The H200 presents a different behaviour. Despite sharing the same FP64:FP32 peak ratio (1/2) as the A100 and offering both higher FP64 throughput and memory bandwidth, the \texttt{IFDDA DP} and \texttt{SP} solver times on the H200 are nearly identical or even higher for SP. This indicates a latency-bound regime. Unlike the \adda case discussed in~\ref{Appendix:GPU}, this behavior cannot originate from CPU--GPU transfers, since \ifdda is fully device-resident and the comparison is purely between double and single precision. A plausible explanation is that, on the H200, the double-precision solver already operates near a practical performance limit set not by bandwidth but by internal latency overheads on the FFT kernels side.

Overall, the results show that, on workstation GPUs with limited FP64 throughput and small memory bandwidth, the choice of precision strongly affects both solver and total time, and significant gains can be obtained by using single precision when iterative convergence permits. On high-bandwidth cluster GPUs, however, the benefits of switching to single precision may decrease or even vanish if the solver approaches an latency-limited regime.

\subsection{GPU vs CPU}
\label{sec:CPUvsGPU}

This section compares the best GPU runtimes obtained on the NVIDIA A100 with the best CPU results measured on the AMD EPYC node. Because the available core counts differ between the two grid sizes, the CPU reference is defined as the fastest configuration yielding the smallest wall-clock time in the CPU benchmarks: 75 cores for $n_x=150$ in the case of \ifdda, 150 cores for $n_x=150$ in the case of \adda, and 50 and 125 cores respectively for $n_x=250$, in order to avoid taking advantage of \adda multi-node implementation.

For $n_x=150$, the best CPU runtime for \adda is \qty{8.3}{s}, whereas the A100 requires \qty{14}{s}. Thus, \adda in OpenCL mode is about 1.7$\times$ slower than the corresponding multi-core CPU execution. In contrast, \ifdda benefits from GPU acceleration, as the double-precision GPU time is 4.1$\times$ faster, and the single-precision time yields a 4.3$\times$ speedup.

At $n_x=250$, \adda becomes 1.8$\times$ slower on the GPU than on the 125-core CPU run, whereas \texttt{IFDDA DP} and \texttt{IFDDA SP} achieve 2.2$\times$ and 2.4$\times$ speedups, respectively.

For the \cmd{ADDA OCL_BLAS} mode on the A100, we observe a 1.3$\times$ speedup relative to the 125-core CPU run (computed for BiCG solver). However, for very large grids (e.g., $n_x=400$), a pure CPU MPI run becomes faster, even though the GPU-based solver itself is 15--20\% faster. In this regime, the performance is limited by the initialization stage, which is not parallelized in the OCL mode and therefore dominates the overall runtime. Introducing MPI parallelism in the OCL initialization stage, or ideally offloading it to the GPU, would remove this limitation, at the expense of a higher memory footprint and a reduced maximum attainable grid size.

Finally, SLURM energy accounting on the AMD node shows that a typical large MPI run of \adda (75~cores) consumes about \qty{700}{kJ}, whereas the single-GPU A100 run requires only \qty{36}{kJ}. Even when the GPU is slower in wall time, its energy efficiency is therefore much higher.

Overall, GPUs provide substantial benefits for fully GPU-resident solvers (\ifdda and \cmd{ADDA OCL_BLAS}), especially when single precision is usable. For \adda, GPU mode is competitive for small and medium grids, but CPU-based MPI remains necessary for the largest cases. Memory capacity is the critical factor for the latter: even modern accelerators offer far less RAM than multi-socket CPU nodes, constraining feasible grid sizes.

\subsection{Memory performance}
From Table~\ref{tab:gpu_RAM}, \ifdda has a larger footprint than \adda, as \texttt{IFDDA SP} and \texttt{IFDDA DP} require about 2 and 4$\times$ the RAM of \adda, respectively. This ratio is higher than what we observe on CPU memory (Section~\ref{sec:memory} reports 2.5$\times$ for a single rank) due to \adda offloading only the FFT data on GPU compared to \ifdda.

\begin{table}[!ht]
\centering
\caption{Peak GPU device memory (\unit{GiB}) for two problem sizes (i.e., $n_x=150$ and $250$).}
\label{tab:gpu_RAM}
\begin{tabular}{lccc}
\toprule
Method & $n_x=150$ & $n_x=250$ \\
\midrule
\adda         & 1.5 & 6.5  \\
\texttt{IFDDA SP}     & 2.9 & 13.4 \\
\texttt{IFDDA DP}     & 5.8 & 28.5 \\
\bottomrule
\end{tabular}
\end{table}

Precision has, in trend, the expected first–order effect, i.e., switching from DP to SP reduces per–element storage by a factor of 2. Thus, when the iterative convergence permits single precision, the 2$\times$ memory reduction can be decisive for feasibility on GPU memory-limited hardware.

Practically, the grid $n_x=250$ \texttt{IFDDA DP} peak (\qty{28.5}{GiB}) requires a \qty{40}{GiB} GPU card (e.g., RTX 6000 Ada), whereas \texttt{IFDDA SP} (\qty{13.4}{GiB}) fits on \qty{16}{GiB} workstations, and \adda (\qty{6.5}{GiB}) remains broadly portable. On the A100 \qty{80}{GiB}, our measurements indicate a feasible maxima around $n_x\!\approx\!350$ for IFDDA DP, $n_x\!\approx\!450$ for \texttt{IFDDA SP}, and $n_x\!\approx\!530$ for \adda. With \adda \cmd{OCL_BLAS} mode, the limit is slightly reduced to $n_x\!\approx\!510$ (for BiCG). Note that \adda provides the \texttt{-opt mem} option, which enforces a memory-conservative execution mode by processing the computational domain one layer at a time along the $x$ direction. This reduces the GPU memory footprint and can therefore slightly increase the maximum feasible grid size. For example, for $n_x=530$, the maximum GPU memory usage decreases from \qty{78.2}{GiB} to \qty{71.7}{GiB}. In addition to reducing memory consumption, this option may also improve runtime stability, as operating close to the GPU memory limit can lead to inefficiencies and performance degradation. The overall performance impact of this trade-off warrants further investigation.

\section{Conclusion}
\label{sec:conclusion}

We presented a unified benchmarking and interoperability software-assisted methodology for the three most widely used implementations of DDA (\ddscat, \adda, and \ifdda). By aligning all physical, numerical, and solver parameters, we demonstrated that the three codes can reach machine-precision agreement on identical test cases. Although applied here to only three solvers, the software package itself is fully general, as any DDA implementation can be integrated into the same cross-verification and benchmarking methodology. This establishes that modern DDA solvers are both numerically mature and internally consistent, and it enables performance comparisons that are decoupled from accuracy differences.

Our CPU benchmarks show that FFT strategy is the dominant factor in performance, as MKL and FFTW substantially outperform legacy GPFA algorithms. Among the tested codes, \adda achieves the best scalability thanks to its 3D--1D FFT decomposition and optimized MPI implementation.

GPU benchmarks confirm that offloading the entire solver phase to the device, as implemented in \ifdda (and in \adda experimental \cmd{OCL_BLAS} mode), provides the strongest acceleration. By contrast, \adda default OpenCL mode offloads only the FFTs while keeping the solver loop on the CPU, making its performance sensitive to CPU--GPU-transfer latency. GPU performance also depends primarily on architectural factors such as memory bandwidth for cluster-type cards, while peak \unit{FLOP/s} can have an impact on workstation cards.

Overall, this study demonstrates that numerical interoperability between DDA codes is not only achievable but also highly informative. Once physical and numerical parameters are harmonized, the remaining performance differences reveal genuine architectural and implementation effects. Our methodology thus enables reproducible, cross-platform verification and provides fair comparisons between independent DDA implementations.

\section*{Acknowledgments}

This work was supported by the Normandy Region (project RADDAERO). Part of the simulations were performed using computing resources of CRIANN (Normandy, France). The authors also thank Bruce Draine and Piotr Flatau for their comments on the DDSCAT code.

\section*{Data availability}
All data and scripts used to produce the benchmark results and figures in this study are publicly available at Zenodo: \url{https://doi.org/10.5281/zenodo.18847515}. To reproduce the software setup, patches must be applied to the exact base versions documented in the repository README.

\appendix
\setcounter{figure}{0}
\setcounter{table}{0}
\section{Minor code modifications}
\label{Appendix:Modifications}

To ensure numerical consistency and fair performance comparison across \ddscat, \adda, and \ifdda, several minor edits were made to \ddscat (version~7.3.4). These modifications, listed below, improve precision handling and solver consistency but do not affect physical results. Complete patch files are available in the Zenodo repository (see Data availability, folder \texttt{patches/}). We have also submitted these files to the \ddscat authors for potential inclusion in the next release. Similar minor changes to \ifdda (during the course of this study) are not listed here, since they were already included in the latest release.

\subsection{Precision of numbers in output}
In order to test floating-point consistency in double precision, output quantities must be printed with 16 significant digits. This required minor modifications to the output routines of \adda and \ddscat.  

\subsection{Modifications of \ddscat}
\paragraph{Precision handling in LDR coefficients}
The literal precision of the Lattice Dispersion Relation (LDR) coefficients was increased to ensure correct rounding in double precision (see~\ref{Appendix:Ldr_derivation}). This eliminates rounding differences observed at the $7$–$8$ significant-digit level between \ddscat and the other implementations.

\paragraph{Precision-consistent in trigonometric routines}
In the cosine–sine integral subroutine, numerical constants were extended to full precision, tolerance thresholds were made precision-dependent, and complex variables were explicitly typed to prevent down-casting. This change does not affect single-precision results, but avoids loss of precision in double ones.

\paragraph{Solver restart policy}
BiCGStab restarts were disabled to align the solver iteration counts and timing with those of \adda and \ifdda, which do not use restarts.

\paragraph{Preconditioner and stopping criterion}
By default, \ddscat applies a left Jacobi preconditioner and a true-residual stopping criterion, resulting in an additional matrix–vector product per iteration.  
For benchmarking equivalence, we disabled preconditioning and used a relative-residual criterion, consistent with the other DDA codes. This change avoids the extra matrix–vector operation and ensures the same linear operator is solved in all cases. Note also that the resulting behavior is the same as that already used for other iterative solvers in \ddscat.

\section{Evaluation of the LDR Coefficients}
\label{Appendix:Ldr_derivation}

The following summarizes the steps required to evaluate the lattice dispersion relation (LDR) coefficients with arbitrary numerical precision. We use the equivalent representations of Eqs.~(3.19) and (A.4) from Ref.~\cite{draine1993beyond},
\begin{equation}
g(\beta,x)
\defeq \sum_{n=-\infty}^{\infty} e^{-x(\beta+n)^2} 
= e^{-x\beta^2}\,\theta_3\left(\ii x \beta,e^{-x}\right)
= \sqrt{\frac{\pi}{x}}\;\theta_3\!\left(\pi\beta,e^{-\pi^2/x}\right),
\label{eq:g_definition}
\end{equation}
where the elliptic theta functions ensure rapid convergence for $x\geq\pi$ and $x\leq\pi$, respectively~\cite{fenton1982} . In particular, they imply
\begin{equation}
g(0,x)=\sqrt{\frac{t}{\pi}}\,g(0,t),\qquad t\defeq\frac{\pi^2}{x}.
\label{eq:g_identity}
\end{equation}

The coefficient $c_1$ is defined by Eq.~(4.2) of Ref.~\cite{draine1993beyond}:
\begin{equation}
c_1=\frac{2}{3}\int_{0}^{\infty}\dd x\left[g^3(0,x)-\left(\frac{\pi}{x}\right)^{3/2}-1\right].
\end{equation}
This integral is split at $x=\pi$, and the change of variables $x\rightarrow t=\pi^2/x$ is applied to the interval $(0,\pi)$. Using Eq.~\eqref{eq:g_identity}, this yields
\begin{equation}
c_1=\frac{2}{3}\int_{\pi}^{\infty}\dd x\left(1+\sqrt{\frac{\pi}{x}}\right)\left[g^3(0,x)-1\right]-2\pi.
\end{equation}

The cubic power of $g$ can be expressed as a lattice sum over all nonzero integer vectors (Eq.~(A.3) of Ref.~\cite{draine1993beyond}),
\begin{equation}
g^3(0,x)-1=\sum_{\mathbf{n}\neq\mathbf{0}} e^{-x|\mathbf{n}|^2}.
\end{equation}
Thus, defining $u=\pi|\mathbf{n}|^2$, one can obtain
\begin{equation}
c_1=\pi\left\{
\frac{2}{3}\sum_{\mathbf{n}\neq\mathbf{0}}
\left[
\frac{e^{-u}}{u}+\sqrt{\frac{\pi}{u}}\,\mathrm{erfc}(\sqrt{u})
\right]
-2
\right\}.
\end{equation}
This can also be written in terms of incomplete gamma functions $\Gamma(1,u)$ and $\Gamma(1/2,u)$, making the whole approach of evaluating the lattice sums similar to that in Ref.~\cite{NIJBOER1958422}.

The second coefficient is defined by Eq.~(4.3) of Ref.~\cite{draine1993beyond}:
\begin{equation}
c_2=\frac{1}{2}\int_{0}^{\infty}\dd x\,g(0,x)\frac{\partial g(0,x)}{\partial x}\left.\frac{\partial^2 g(\beta,x)}{\partial \beta^2}\right|_{\beta=0}.
\label{eq:c2_definition}
\end{equation}

Derivatives are evaluated using Eqs.~\eqref{eq:g_definition} and \eqref{eq:g_identity}, yielding
\begin{equation}
\frac{\partial^2 g(\beta,x)}{\partial \beta^2}=-2x\left[g(\beta,x)+2x\frac{\partial g(\beta,x)}{\partial x}\right],
\end{equation}
\begin{equation}
2x\frac{\partial g(0,x)}{\partial x}=-\sqrt{\frac{t}{\pi}}\left[g(0,t)+2t\frac{\partial g(0,t)}{\partial t}\right].
\end{equation}
Using these identities, separating the integral at $x=\pi$, and transforming $g$ back into the sum, Eq.~\eqref{eq:c2_definition} can be rewritten as
\begin{equation}
c_2
=\int_{\pi}^{\infty} dx
\left(1+\sqrt{\frac{\pi}{x}}\right)
\sum_{\mathbf{n}\neq\mathbf{0}}
e^{-x|\mathbf{n}|^2}
\left[
x n_z^2 - 2x^2 n_y^2 n_z^2
\right].
\label{eq:c2_intermediate}
\end{equation}

The integration of each term in the sum leads to incomplete gamma functions $\Gamma(3/2,u)$, $\Gamma(2,u)$, $\Gamma(5/2,u)$, and $\Gamma(3,u)$, which admit closed-form expressions in terms of $\exp(-u)$ and $\mathrm{erfc}(\sqrt{u})$ ~\cite{dlmf_gamma_8_4}. Due to the cubic symmetry, $n_z^2$ can be replaced by $|\mathbf{n}|^2/3$, while $n_y^2 n_z^2$ -- by $|\mathbf{n}|^4 v/9$, where
\begin{equation}
v \defeq
\frac{3\left(n_x^2 n_y^2+n_y^2 n_z^2+n_z^2 n_x^2\right)}{|\mathbf{n}|^4},
\qquad 0\le v\le 1 .
\label{eq:v_definition}
\end{equation}

Finally, one obtains
\begin{equation}
c_2=
\frac{\pi}{3}
\sum_{\mathbf{n}\neq\mathbf{0}}
\left\{
\frac{e^{-u}}{u}
\left[
1+2u-\frac{v}{3}\left(4+7u+4u^2\right)
\right]
+
\left(\frac{1-v}{2}\right)
\sqrt{\frac{\pi}{u}}\,
\mathrm{erfc}(\sqrt{u})
\right\},
\qquad
\end{equation}
with $u=\pi|\mathbf{n}|^2$.

The third coefficient satisfies the exact identity
\begin{equation}
c_1-4c_2+2c_3=
-\frac{4}{3}\int_0^\infty \dd x\frac{\partial}{\partial x}
\left\{x\left[g^3(0,x)-\left(\frac{\pi}{x}\right)^{3/2}-1\right]\right\}
=0,
\label{eq:c_identity}
\end{equation}
which follows directly from their definitions in Ref.~\cite{draine1993beyond} and the fact that the function under the derivative vanishes at both ends of the integration domain. Therefore, $c_3$ does not need to be evaluated independently. This identity has also been indirectly proven in Appendix A2 of Ref.~\cite{Mackowski2002}.

When evaluated with sufficient precision to ensure accuracy up to quad (128-bit) arithmetic (using truncation at $|\mathbf{n}|\leq 5$ for $c_1$ and $|\mathbf{n}|\leq 5.2$ for $c_2$), the final LDR coefficients (in CGS units) are
\begin{align}
b_1 &=-\frac{c_1}{\pi}=  1.8915316529870796511106114030718259, \\
b_2 &=-\frac{c_2}{\pi}= -0.16484691508771947306079362778185226, \\
b_3 &=\frac{3c_2+c_3}{\pi}= 1.7700004019321371908592738404451742.
\end{align}
Importantly, we use the sign convention employed in the ADDA code~\cite{yurkin_ADDA_2011}, opposite to the original paper~\cite{draine1993beyond}, to facilitate comparison with other DDA formulations~\cite{yurkin_review_2007}. When expressed in SI units, the coefficients $b_1$ and $b_3$ should be additionally divided by $4\pi$. These values have been used in \adda since 2022 (commit \texttt{887a63c}) and they satisfy
\begin{equation}
b_1-10b_2-2b_3=0
\end{equation}
due to Eq.~\eqref{eq:c_identity}.

As a final note, the above approach may also be used to calculate LDR coefficients for rectangular cuboid lattices~\cite{gutkowicz2004,SMUNEV2015Rect}.

\section{Equivalence of parameters}
\label{Appendix:equivalence}

Performing equivalent simulations with \adda, \ifdda, and \ddscat requires matching the command-line parameters level by level, following the accuracy hierarchy introduced in Table~\ref{tab:parameter_accuracy}. In practice, one starts from the executables \texttt{adda} and \texttt{ifdda}, and the Python wrapper \texttt{ddscatcli}~\cite{Argentin_ddscatcli_2025}, then progressively enforces the equivalences listed in Table~\ref{tab:cli_equivalences_levels} ignoring all lines that are not relevant (not used by any of the codes). The number of matching digits is then determined by the first non-matching line. If all lines in the table are matched, the machine-precision level is reached (10--15 digits). 

Table~\ref{tab:cli_equivalences_levels} is designed to show the correspondence between parameters, rather than to provide a list of specific numerical values that would enforce strict equivalence across codes, since achieving this requires additional care discussed below. We also provide a set of cross-code equivalent command-line examples in the Python package~\cite{argentin_clement_2026_dda-bench} (in the file \cmd{DDA_commands}) as concrete working examples. Note that even with the \texttt{ddscatcli} wrapper, the user must still create or modify the input file for the refractive index if it is not available in the \ddscat examples folder.

A few conventions are essential for interpreting  
Table~\ref{tab:cli_equivalences_levels}. First, particle size and shape definitions differ across codes. In particular, \ddscat defines the target size through an equivalent radius $a_{\mathrm{eff}}$. For a given volume $V$, this corresponds to
\begin{equation}
    a_{\mathrm{eff}}=\left(\frac{3V}{4\pi}\right)^{1/3}.
\end{equation}
For shapes affected by volume correction, a direct one-to-one mapping of the nominal radius or diameter does not preserve the exact physical volume across codes. Consequently, the particle size must be adjusted in either \ifdda or \ddscat to ensure volume equivalence (\adda can use \cmd{-no_vol_cor} to switch). For inhomogeneous materials (e.g., coated spheres), arbitrary shapes in \ifdda (which are treated as box-embedded), and for the diagonally anisotropic materials (\ifdda requires CM or RR polarizability), the maximum achievable agreement is limited to the $\eta$-level, even when all other parameters are matched. Indeed, in these cases, the linear systems cannot be made strictly equivalent across codes (see Sec.~\ref{sec:linear_systems}). The bisphere is treated as an inhomogeneous material in both \ifdda and \ddscat and therefore requires the corresponding material specification. In this work, identical dielectric functions are assigned to all components to represent a homogeneous bisphere and achieve full digit agreement between the codes. For coated spheres, \ddscat further requires three dielectric components when using \cmd{-SHPAR 1}, following the anisotropic-shell configuration. Note that for arbitrary shapes, \ddscat and \adda share a common geometry file format (original to \ddscat, based on integer voxel positions), enabling direct reuse between the two codes. \ifdda, by contrast, employs a distinct format based on real voxel coordinates. Additionally, in contrast to \ddscat and \adda, \ifdda require explicit grid size in this file, in addition to the corresponding input parameter used for memory allocation for both \ifdda and \ddscat (\cmd{-nnnr} and  \cmd{-MEM_ALLOW}). Moreover, this grid parameter corresponds to the maximum particle extent in \ifdda, whereas \adda expects one along the $x$ direction (i.e.\ $n_x$). Finally, as discussed in Sec.~\ref{sec:default_parameters}, the physical units used for wavelength and particle size are arbitrary as long as the relevant ratios are preserved and the outputs are rescaled consistently.

In addition, the incident propagation direction and the target orientation are treated differently between the codes (although the same unit of degrees is assumed for input angles). Both \ddscat and \adda build the voxel grid in a target-fixed reference frame (Target Frame, TF), while the incident beam is described in the laboratory frame (Lab Frame, LF). To rotate the particle without rebuilding the grid, both codes use a passive approach, i.e., instead of actively rotating the dipole coordinates, they rotate the incident propagation and polarization vectors (and scattering directions) into the TF. Thus, rotating the target is completely equivalent to that of the incident beam -- the corresponding matrices are inverse of each other.

Despite this common strategy, the two codes differ by conventions and default axis choices. \adda employs a passive Euler rotation matrix in the $zyz$ convention and the default $\hb{k}_0=\hb{z}$, while \ddscat --- $xzx$ convention~\cite{yurkin2023orientation} and hard-coded $\hb{k}_0=\hb{x}$.  Here, $\hb{k}_0$ denotes the incident propagation in the LF to discriminate it from $\hb{k}$ in the TF that is used in the DDA equations. Moreover, for some shapes (e.g.\ ellipsoids or \cmd{SPHERES_N}), \ddscat allows one to interchange or rotate the target axes (denoted by $\hb{a}_i$), which corresponds to preliminary active rotation of the particle --- we do not investigate this feature and assume the default $\hb{a}_i$.

While any combination of  $\hb{k}_0$ and Euler angles in \adda can be matched by \ddscat, one-to-one correspondence takes place only if $\hb{k}_0$ is fixed. Most naturally for the latter is to use $\hb{k}_0=\hb{x}$, corresponding to  \texttt{-prop 1 0 0} in \adda, which allows us to disentangle equivalence conditions for propagation direction and orientation. The latter then boils down to equality of Euler-rotation matrices with different conventions
\begin{equation}
    \mathbf{R}_{zyz}(\alpha,\beta,\gamma)=\mathbf{R}_{xzx}(\Phi,\Theta,\beta_\mathrm{D}),
    \label{rotations}
\end{equation}
where we denote the third Euler angle in \ddscat as $\beta_\mathrm{D}$ to discriminate $\beta$ used in its manual from that of \adda. As expected, default orientations (all zero angles) correspond to each other, but otherwise the transformation of angles expressions involve direct and inverse trigonometric functions.

By contrast, \ifdda applies an active rotation, i.e., the voxel coordinates are defined in the LF and then rotated to be tested against the particle shape in the TF. This approach is physically equivalent to the passive formulation, but it does not reuse the same voxel grid for different orientations. Effectively, the DDA equations are solved in the LF rather than in the TF. Thus, exact agreement between \ifdda and other codes is only expected for rotations that map the grid lattice onto itself, i.e., for all angles being multiples of 90$^\circ$. Otherwise, the agreement is limited to method error.

The propagation direction in \ifdda can be freely specified as in \adda, but using spherical angles instead of Cartesian components of $\hb{k}_0$. Thus,  \cmd{-prop 90 0} in \ifdda corresponds to $\hb{k}_0=\hb{x}$ in \ddscat, in which case, the rotation matrix $\mathbf{R}_{zxz}(\psi,\theta,\varphi)$ should be equal to the ones in Eq.~\ref{rotations} (all three using different conventions). Angles $\theta,\varphi$ are as used in \ifdda manual; they should not be confused with angles defining incident or scattering directions.

In the LF, \ddscat uses $\hb{y}$ and $\hb{z}$ as basic polarization vectors, while \adda and \ifdda employ $\hb{y}$ and $\hb{x}$ with the default $\hb{k}_0=\hb{z}$. When the propagation vector is rotated into $\hb{x}$, the latter two polarizations become $\hb{y}$ and $-\hb{z}$, i.e., the same as in \ddscat apart from the minus. This minus sign has no effect for most of the simulated quantities, except for internal fields. As discussed below, $\hb{y}$ defines the default scattering plane in both \adda and \ddscat for both of the discussed directions of the incident field. Thus, we denote it as $\parallel$ to this plane and the complementary polarization as $\perp$. This designation remains meaningful for any $\hb{k}_0$.

Although \ddscat is the most flexible code in specifying the single incident polarization by a full complex vector (which can match any setting of \ifdda and \adda), the second polarization is automatically obtained from the first (by default, it is $\hb{y}\rightarrow\hb{z}$). That means that \cmd{-IORTH 2} for \ddscat would use the same two polarizations as \adda, but the second one will have an inverse sign. Alternatively, one may use \cmd{-IORTH 2 -POL_E01 "(0,0) (0,0) (-1,0)"} to use exactly the same polarizations, but in reverse order.

\ifdda supports circular polarizations, as well as arbitrary linear ones. The latter is controlled by a single real parameter \texttt{pola} (with $0 \le \texttt{pola} \le 1$), which sets the relative weight of two basic polarizations. This can be mapped to \ddscat by specifying the complex incident polarization unit vector $\hb{e}_{01}$ as
\begin{equation}
    \hb{e}_{01} = (0,s,-p ), \, 
    s \defeq \sqrt{1-\texttt{pola}},\, p\defeq \sqrt{\texttt{pola}},
\end{equation}

\adda is designed mostly for simulating two incident polarizations to obtain the Mueller matrix. To simulate a single one (and not waste computational resources), one can trick \adda into considering the scatterer symmetric, but then produced Mueller matrix will generally be incorrect (and better discarded). Thus, addition of \cmd{-sym enf -scat_matr none} is a workaround to simulate only $\parallel$ polarization. To rotate this single linear polarization, there are two possible workarounds. First, it can be emulated by additional rotation of the particle. However, no simple expressions exist for arbitrary particle orientation and $\hb{k}_0$. Still, a few special case can be named: for $\hb{k}_0=\hb{z}$, one only need to increment Euler angle $\gamma\rightarrow\gamma+\arcsin{p}$, while for $\hb{k}_0=\hb{x}$ and default particle orientation we have $\alpha =-\pi/2$, $\beta =\arcsin{p}$, $\gamma =\pi/2$. The second option is through the general Bessel beams implemented in \adda, by setting both order and half-cone angle to zero: \cmd{-beam besselM 0 0 0 0 <p> <s>}, although it is incompatible with calculation of radiation forces. Each of the workarounds should be combined with the above symmetry trick, otherwise the behavior will be similar to \cmd{-IORTH 2} in \ddscat.

For circular polarizations, we assume the definition of \ifdda, in agreement with Ref.~\cite{bohren1983absorption}, specifically, $\hb{e}_\mathrm{R,L} \defeq (\hb{y} \pm \ii \hb{x})/\sqrt{2}$ for  right- and left-circular polarizations, respectively, of a beam propagating along $\hb{z}$. That is essentially opposite to the ones given in \ddscat manual, which are equivalent to $\hb{e}_\mathrm{L}$ and $\ii\hb{e}_\mathrm{R}$, respectively, after rotation of $\hb{k}_0$. In \adda, such polarizations can be emulated through the Bessel beam, for instance, \cmd{-beam besselM 0 0 0 0 0 <is2> 0 0 <}$\pm$\cmd{is2> 0} for $\hb{e}_\mathrm{R,L}$, respectively, where $\texttt{is2}=1/\sqrt{2}$. Same as above, such emulation is not suitable for radiation forces. Moreover, computation of two polarizations with ADDA is completely meaningless in this case, since those two polarizations will be equivalent (in contrast to \cmd{-IORTH 2} in \ddscat). Finally, the QMR solver (as well as QMR2 and BiCG) breaks down for circular polarizations, since the pseudo--norm of incident field vanishes \cite{ADDAmanual_2020}; this can be resolved by switching to BiCGStab.

For the isotropized LDR polarizability, \adda and \ifdda use different conventions. \adda averages the term $S$ only over incident polarizations keeping the propagation direction fixed (so the result depends on the latter, see Eq.~(46) in Ref.~\cite{yurkin_review_2023}), whereas \ifdda performs full averaging leading to $\langle S\rangle=1/5$. Therefore, the incident propagation angles must be chosen carefully to reproduce the same value of $S$. An example is 
$\hb{k} = \left( \sqrt{(5-\sqrt{5})/10},0,\sqrt{(5+\sqrt{5})/10} \right)$,
which corresponds to angle $\theta=(1/2)\arctan{2}$ in \ifdda.

Several additional numerical aspects are worth noting. The option \texttt{-scat fin} of \adda affects the value of $C_\mathrm{abs}$ only for some combination of other parameters (like absorbing particles with the LDR polarizability). By contrast, enabling \cmd{igt_so} reduces the accuracy of $C_{\mathrm{ext}}$ to 1--2 significant digits compared to full IGT, as for other variations of formulations. However, most differences in the formulation class are $\mathcal{O}[(kd)^2]$, which decreases with particle size (for fixed discretization). The notable exception is the interaction formulations, which remain significant even for particles much smaller than the wavelength. Finally, the equivalence of the BiCGStab solver is problem dependent, due to the last residual stopping criterion.

\begin{landscape}
\begin{table}[!ht]
\centering
\tiny
\begin{threeparttable}
\caption{Command-line equivalences between \ifdda, \adda, and \ddscat, organized by accuracy-critical levels.}
\label{tab:cli_equivalences_levels}
\begin{tabular}{p{1.5cm} p{3.2cm} p{4.5cm} p{5.0cm} p{6.0cm} p{0.7cm}}
\toprule
Level & Parameter & \ifdda & \adda & \ddscat & Digits \\
\midrule

\multirow{8}{*}{Shape}
& Sphere
& \cmd{-object sphere <R>}
& \cmd{-shape sphere -size <D>}
& \cmd{-CSHAPE ELLIPSOID -SHPAR "<Nx> <Nx> <Nx>" -AEFF "<aeff> <aeff> 1 'LIN'"}
& \multirow{8}{*}{0} \\

& Cube
& \cmd{-object cube <D>}
& \cmd{-shape box 1 1 -size <D>}
& \cmd{-CSHAPE RCTGLPRSM -SHPAR "<Nx> <Nx> <Nx>" -AEFF "..."}
&  \\

& Cuboid
& \cmd{-object cuboid1 <Dx> <Dy> <Dz>}
& \cmd{-shape box <Dy/Dx> <Dz/Dx> -size <Dx>}
& \cmd{-CSHAPE RCTGLPRSM -SHPAR "<Nx> <Ny> <Nz>" -AEFF "..."}
&  \\

& Cylinder
& \cmd{-object cylinder <R> <h>}
& \cmd{-shape cylinder <h/D> -size <D>}
& \cmd{-CHSAPE CYLNDRPBC -SHPAR "<Nz> <Nx> 3 0 0 1" -AEFF "..."}
&  \\

& Ellipsoid
& \cmd{-object ellipsoid <Rx> <Ry> <Rz>}
& \cmd{-shape ellipsoid <y/x> <z/x> -size <Dx>}
& \cmd{-CHSAPE ELLIPSOID -SHPAR "<Nx> <Ny> <Nz>" -AEFF "..."}
&  \\

& Bisphere
& \cmd{-object nspheres 2 <R> <R>}
& \cmd{-shape bisphere 1 -size <Dx>}
& \cmd{-CSHAPE ELLIPSO_2 -SHPAR "<Nx> <Ny> <Nz>" -AEFF "..."}\tnote{a}
&  \\

& Coated sphere
& \cmd{-object concentricsphere 2 <Rin> <Rout>}
& \cmd{-shape coated <Din/D> -size <D>}
& \cmd{-CSHAPE ONIONSHEL -SHPAR "1 <D> <Din/D>" -AEFF "..."}
&  \\

& Arbitrary
& \cmd{-object arbitrary <filename>}
& \cmd{-shape read <filename> -size <Dx>}
& \cmd{-CSHAPE FROM_FILE -AEFF "..."}
& \\

\multirow{1}{*}{Particle}
& Orientation
& \cmd{-orient <phi> <theta> <psi>}
& \cmd{-orient <alpha> <beta> <gamma>}
& \cmd{-ROT_BETA "<beta> <beta> 1" -ROT_THETA "<Theta> <Theta> 1" -ROT_PHI "<Phi> <Phi> 1"}
&  \\

\midrule

\multirow{3}{*}{Material}
& Isotropic
& \cmd{-epsmulti <eps_r> <eps_i>}
& \cmd{-m <m_r> <m_i>}
& \cmd{-DIEL "filename"}
& \multirow{3}{*}{0} \\

& Diagonal anisotropic
& \cmd{-trope <eps1_r> <eps1_i> ... <eps3_r> <eps3_i>}
& \cmd{-anisotr -m <m1_r> <m1_i> <m2_r> <m2_i> <m3_r> <m3_i>}
& \cmd{-NCOMP 3 --diels "filename1" "filename2" "filename3"}
&  \\

& Inhomogeneous
& \cmd{-epsmulti <...>} (per material)
& \cmd{-m <...>} (per material)
& \cmd{-NCOMP <Ncomp> --diels "..."} (per material)
& \\

\midrule

\multirow{8}{*}{Incident beam}
& Linear ($\parallel$ or $\hb{y}$)
& \cmd{-beam pwavelinear 0 0}
& \cmd{-beam plane -sym enf -scat_matr none}
& \cmd{-IORTH 1 -POL_E01 "(0,0) (1,0) (0,0)"}
& \multirow{8}{*}{0} \\

& Linear ($\perp$ or $\hb{x}$)
& \cmd{-beam pwavelinear 1 0}
& natively unsupported\tnote{b}
& \cmd{-IORTH 1 -POL_E01 "(0,0) (0,0) (-1,0)"}
&  \\

& Linear (arbitrary)
& \cmd{-beam pwavelinear <pola> 0}
& natively unsupported\tnote{b}
& \cmd{-IORTH 1 -POL_E01 "(0,0) (<s>,0) (-<p>,0)"}
&  \\

& Two linear polarizations
& ---
& \cmd{-beam plane}
& \cmd{-IORTH 2 -POL_E01 "(0,0) (1,0) (0,0)"}
&  \\

& Circular (right)
&\cmd{-beam pwavecircular 1}
& natively unsupported\tnote{b}
& \cmd{-IORTH 1 -POL_E01 "(0,0) (1,0) (0,-1)"}
&  \\

& Circular (left)
& \cmd{-beam pwavecircular -1}
& natively unsupported\tnote{b}
& \cmd{-IORTH 1 -POL_E01 "(0,0) (1,0) (0,1)"}
&  \\

& Propagation direction
& \cmd{-prop <theta> <phi>}
& \cmd{-prop <x> <y> <z>}
& ---
&  \\

& Wavelength
& \cmd{-lambda <value>}
& \cmd{-lambda <value>}
& \cmd{-WAVELENGTHS "<value> <value> 1 'LIN'"}
& \\

\midrule

\multirow{1}{*}{Grid}
& Grid size
& \cmd{-nnnr <N>}
& \cmd{-grid <N>}
& \cmd{-MEM_ALLOW "<Nx> <Ny> <Nz>" -SHPAR "..."}\tnote{c}
& 1--2 \\

\midrule

\multirow{10}{*}{Formulation}
& Clausius-Mossotti
& \cmd{-polarizability CM}
& \cmd{-pol cm}
& ---
& \multirow{10}{*}{1--2} \\

& Radiative reaction
& \cmd{-polarizability RR}
& \cmd{-pol rrc}
& ---
& \\

& Digitized Green's function
& \cmd{-polarizability GB}
& \cmd{-pol dgf}
& ---
& \\

& Lakhtakia
& \cmd{-polarizability LA}
& \cmd{-pol lak}
& ---
&  \\

& LDR (standard)
& \cmd{-polarizability LS}
& \cmd{-pol ldr}
& \cmd{-CALPHA LATTDR}
& \\

& LDR (isotropized)
& \cmd{-polarizability LR}
& \cmd{-pol ldr avgpol}
& ---
& \\

& Integrated Green's tensor
& \cmd{-igt <value>}
& \cmd{-int igt <value>}
& ---
& \\

& Filtered Green
& \cmd{-polarizability FG}
& \cmd{-pol fcd -int fcd}
& \cmd{-CALPHA FLTRCD}
& \\

& POI
& \cmd{-igt 0}
& \cmd{-int poi}
& ---
& \\

& Scattering formula
& ---
& \cmd{-scat dr}
& ---
& \\

\midrule

\multirow{6}{*}{Solver}
& QMR
& \cmd{-methodeit QMRCLA}
& \cmd{-iter qmr}
& \cmd{-CMDSOL QMRCCG}
& \multirow{6}{*}{$\eta$} \\

& BiCGStab
& \cmd{-methodeit BICGSTAB}
& \cmd{-iter bicgstab}
& \cmd{-CMDSOL PBCGST}
&  \\

& GPBiCG
& \cmd{-methodeit GPBICG1}
& ---
& \cmd{-CMDSOL GPBICG}
& \\

& BCGS2
& ---
& \cmd{-iter bcgs2}
& \cmd{-CMDSOL PBCGS2}
&  \\

& Threshold $\eta$
& \cmd{-tolinit 1.d-<n>}
& \cmd{-eps <n>}
& \cmd{-TOL 1.d-<n>}
&  \\

& Initial guess
& \cmd{-ninitest 0}
& \cmd{-init_field zero}
& ---
&  \\

\bottomrule
\end{tabular}
\begin{tablenotes}
\item[a] Here, \texttt{-SHPAR} defines the grid size of one sphere.
\item[b] Not directly available in \adda; see the text for available workarounds.
\item[c] \texttt{-SHPAR} option defines both geometry of the shape (aspect ratios and, sometimes, the main axes) and the level of the discretization ($n_x$, etc.). The accuracy associated with grid is relevant if only the discretization parameters are varied keeping all aspect ratios fixed.
\end{tablenotes}
\end{threeparttable}
\end{table}
\end{landscape}

Table~\ref{tab:cli_equivalences_quantities}, by contrast, describes equivalent command-line parameters to produce additional scattering quantities. They can be optionally added on top of the free parameters listed in Table~\ref{tab:cli_equivalences_levels}, as they do not affect other output of the codes.
When these additional combinations are used exactly as listed, these quantities inherit the accuracy level of the corresponding configuration listed in Table~\ref{tab:cli_equivalences_levels}. Otherwise, the expected accuracy level drops to 0, as it may incur comparison of different types of fields or different components of forces. 

\adda and \ifdda compute the radiation force (and torque in \ifdda) exerted on each dipole~\cite{Chaumet2000force}, whereas \ddscat provides only total integrated values obtained from numerical integration over scattering directions~\cite{draine1996torque}. Moreover, in \ddscat the radiation force $\mathbf{Q}_\mathrm{pr}$ must be post-processed using momentum conservation (i.e.\ $Q_\mathrm{ext}\hb{k} - Q_\mathrm{sca}\,\mathbf{g}$, where $\mathbf{g}$ is the asymmetry parameter). As a result, while machine-precision agreement is achievable between \adda and \ifdda, comparisons involving \ddscat are typically limited to about 4--5 significant digits due to numerical integration errors, even when using millions of scattering directions. Moreover, the above relation is valid only for plane incident waves, although the same limitation is currently present in \adda for dipole-wise calculation of forces.

To compare radiation forces between \ifdda and \adda, additional post-processing is required, since \ifdda outputs the total time-averaged force $\mathbf{F}_{\mathrm{rad}}$ (in \unit{N}), while \adda outputs the radiation-pressure cross section $\mathbf{C}_{\mathrm{pr}}$ (in \unit{nm^2} in our case) and the corresponding efficiency $\mathbf{Q}_{\mathrm{pr}}$. The cross section and the total force are related by
\begin{equation}
    \mathbf{F}_{\mathrm{rad}}
    =
    \mathbf{C}_{\mathrm{pr}}
    \, \left|\mathbf{E}_0\right|^2 \frac{\varepsilon_0}{2}.
\end{equation}

For the radiation torques, the process is similar since \ifdda outputs the total time-averaged torque $\mathbf{\Gamma}_{\mathrm{rad}}$ (in \unit{N.m}), while \ddscat outputs the radiation-torque efficiency $\mathbf{Q}_\mathrm{trq}$. They are related by
\begin{equation}
    \mathbf{\Gamma}_{\mathrm{rad}}
    =
    \mathbf{Q}_{\mathrm{trq}}\pi a_\mathrm{eff}^2
    \, \left|\mathbf{E}_0\right|^2 \frac{\varepsilon_0}{2k}.
\end{equation}

Regarding the internal field, \ifdda and \ddscat compute the macroscopic field by multiplying the local/excited field by, respectively, radiative-reaction correction and the Clausius--Mossotti factor, whereas \adda rigorously computes the internal field according to the selected polarizability (see Eq.~(\ref{eq:Emicro_to_Emacro})). Therefore, machine-precision agreement is achievable only for these two polarizabilities (and not between all three codes); for other choices, the expected level of agreement drops to that obtained when the polarizations are not matched (1--2 digits). Moreover, \ddscat does not support the Clausius--Mossotti polarizability (see Table~\ref{tab:features_comparison}), so full machine precision is not achievable for it as well. One can, in principle, compare instead voxel polarizations between \ddscat and \adda, but this requires additional efforts: the source code of \ddscat need first to be modified to save the internal polarizations multiplied by the voxel volume $d^3$ (to match the \adda convention).

Finally, it is also possible to compare Mueller matrices between \adda and \ddscat. In the latter code (with \texttt{-CMDFRM LFRAME}), the scattering direction is parameterized by $(\theta,\phi)$, where $\phi=0$ corresponds to the reference scattering plane coinciding with the $(\hb{x}_\mathrm{LF},\hb{y}_\mathrm{LF})$ plane. Note that these $\theta$ and $\phi$ are distinct from the particle-orientation angles $\Theta$ and $\Phi$). In \adda, the scattering plane is also set by the incident basis vectors and corresponds to the $(\hb{k},\hb{e}_Y)$ plane. Therefore, once these vectors are mapped between the two formulations (see above), both codes share the same scattering-plane convention, enabling direct comparisons of Mueller matrices. 

Comparison of the amplitude scattering matrices would be more complicated, since \ddscat defines it with respect to fixed incident polarizations~\cite{draine_discrete_1988} rather than to the ones rotated together with the scattering plane~\cite{bohren1983absorption}. Moreover, \ddscat does not produce the original complex elements of this matrix, but rather real quadratic combinations that are further used to compute the Mueller matrix.

\begin{table}[!ht]
\centering
\tiny
\begin{threeparttable}
\caption{Command-line equivalences between \ifdda, \adda, and \ddscat, defining the calculated output quantities.}
\label{tab:cli_equivalences_quantities}
\begin{tabular}{p{2.5cm} p{2.5cm} p{2.5cm} p{3.5cm}}
\toprule
Parameter & \ifdda & \adda & \ddscat \\
\midrule
Field
& \cmd{-near_field 0 1 0 -save_data 2}
& \cmd{-store_int_field}
& \cmd{-NRFLD 1} \\

Polarization
& ---
& \cmd{-store_dip_pol}
& \cmd{-NRFLD 1} \\

Force
& \cmd{-nforce 1 1 0 0}
& \cmd{-Cpr}
& \cmd{-CMDTRQ DOTORQ} \\

Torque
& \cmd{-nforce 1 1 1 1}
& ---
& \cmd{-CMDTRQ DOTORQ} \\

Mueller matrix\tnote{a}
& ---
& \cmd{-scat_matr muel -ntheta 180}
& \cmd{-IORTH 2 -CMDFRM LFRAME -NPLANES 1 -PLANE "0 0 180 1" -NSMELTS 16 -SMELTS_LIST "11 12 13 14 21 22 23 24 31 32 33 34 41 42 43 44"} \\

\bottomrule
\end{tabular}
\begin{tablenotes}
\footnotesize
\item[a] Corresponds to the default \adda options for a single scattering plane with particle symmetric with respect to the $xz$-plane.
\end{tablenotes}
\end{threeparttable}
\end{table}

Table~\ref{tab:matching_digits} provides an example of the matching digits obtained in this work (see the command-lines in Sec.~\ref{sec:comparison}). Note that \adda reports the extinction cross section $C_{\mathrm{ext}}$ in \unit{nm^2} (when \unit{nm} is assumed for input quantities), so the \ifdda values, originally expressed in SI units (\unit{m^2}), were converted to \unit{nm^2} before comparison. By contrast, \ddscat is compared only through the dimensionless efficiency $Q_{\mathrm{ext}}$.

\begin{table}[!ht]
\centering
\footnotesize
\caption{Matching digits for $C_{\mathrm{ext}}$ and $Q_{\mathrm{ext}}$ between \adda, \ifdda, and \ddscat.}
\label{tab:matching_digits}
\begin{tabular}{lccc}
\toprule
Grid size & Code & $C_{\mathrm{ext}}$ [\unit{nm^2}] & $Q_{\mathrm{ext}}$ \\
\midrule
$n_x=150$ & \adda   & \textbf{24729875.277}\,382743 & \textbf{3.58901283529}\,54913 \\
          & \ifdda  & \textbf{24729875.277}\,422981 & -- \\
          & \ddscat & -- & \textbf{3.58901283529}\,78903 \\
\midrule
$n_x=250$ & \adda   & \textbf{24730049.76}\,7323192 & \textbf{3.5890381587}\,809035 \\
          & \ifdda  & \textbf{24730049.76}\,6945954 & -- \\
          & \ddscat & -- & \textbf{3.5890381587}\,500482 \\
\bottomrule
\end{tabular}
\end{table}

\section{MPI memory scaling model}
\label{Appendix:mpi_memory}

In MPI programs, each rank maintains its own process space and communication buffers, which leads to a node-level memory overhead that scales both linearly and quadratically with the number of ranks $P$. We model the total resident set size (RSS) as
\begin{equation}
    \mathrm{RSS}(N, P) \approx D_{\text{ADDA}}(N) + C_{\text{RSS}} P + \alpha_{\text{RSS}} P^{2},
\end{equation}
where $D_{\text{ADDA}}(N)$ is the application-dependent working set predicted by Eq.~(4) of the \adda manual~\cite{ADDAmanual_2020}, $C_{\text{RSS}}$ is the resident portion of the constant per-rank libraries runtime environment used, and $\alpha_{\text{RSS}}$ quantifies the memory usage of MPI communication pools. Preliminary fits on our cluster suggest $C_{\text{RSS}} \approx \qty{20}{MiB}$ and $\alpha_{\text{RSS}} \approx \qty{0.3}{MiB}$, but they are expected to vary depending on hardware and MPI implementation.

The quadratic term is most pronounced in single-node runs with many ranks; for multi-node configurations with fewer ranks per node, the growth approaches linearity. At large grid sizes, $D_{\text{ADDA}}(N)$ dominates and MPI overhead becomes negligible compared to the overall working set. This seems to correspond to the natural design goal of MPI implementations, that most of the node memory should remain accessible to the application itself.

\section{Fully offloaded ADDA solver (\cmd{OCL_BLAS})}
\label{Appendix:GPU}

\begin{figure}[!hb]
    \centering
    \includegraphics[width=\linewidth]{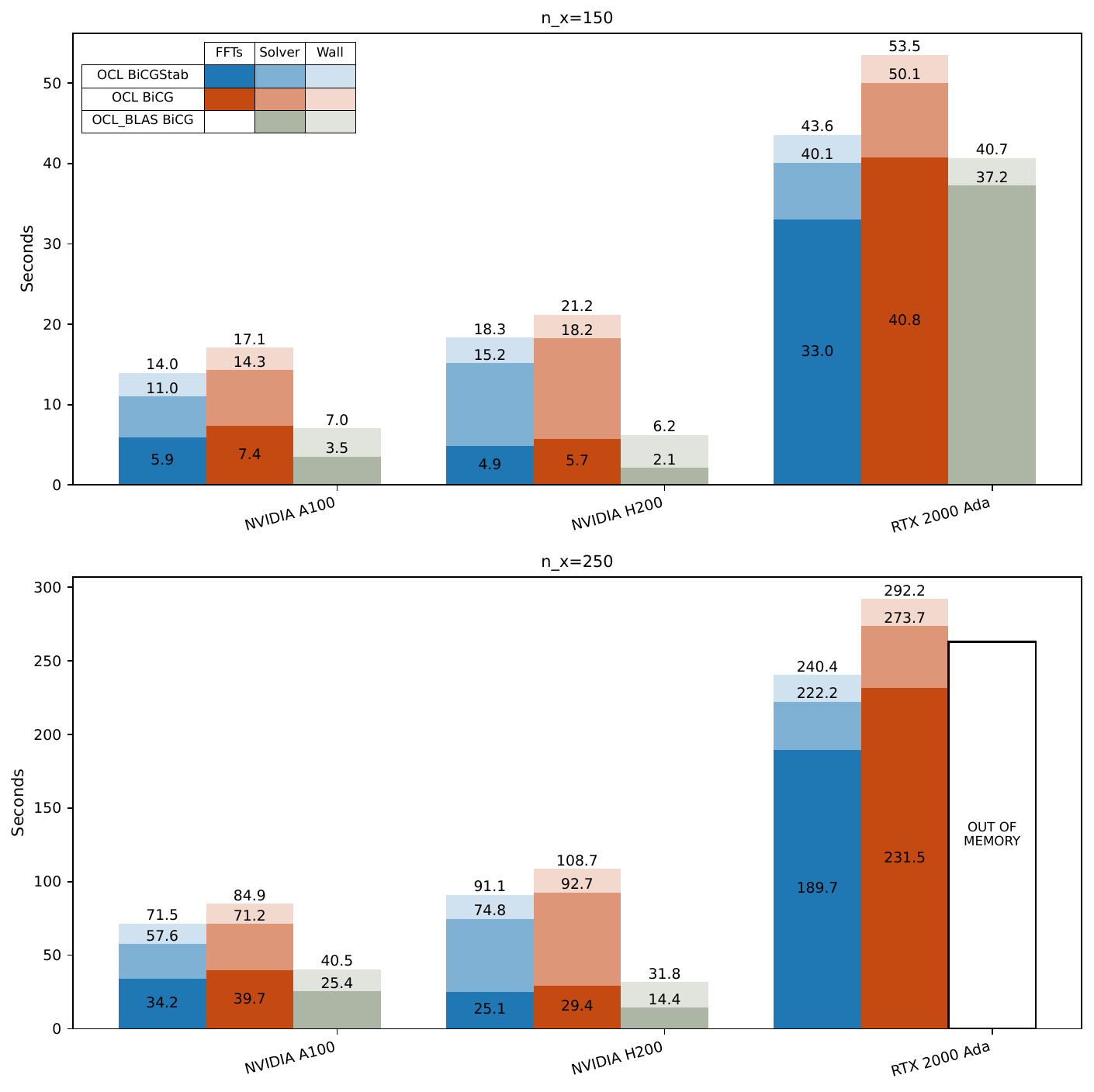}
    \caption{
    GPU timings for \adda and \cmd{ADDA OCL_BLAS} mode. Bars are grouped by mode and solver type and stacked to show FFT time, solver time, and 1-core wall-time. Grid size $n_x=150$ (top row); $n_x=250$ (bottom row). Blue-column results (BiCGStab) are the same as in Fig.~\ref{fig:fig3_dda_runtime_gpu}.
    }
    \label{fig:Appendix_fig1_ocl_blas}
\end{figure}

Figure~\ref{fig:Appendix_fig1_ocl_blas} reports the FFTs time, solver time, and single-core wall-clock time on the A100, H200, and RTX 2000 Ada GPUs, for both $n_x=150$ and $n_x=250$ grids, comparing the \cmd{OCL_BLAS} mode of \adda (BiCG solver) with the default OpenCL backend using the same BiCG solver. This ensures a fair comparison with \cmd{OCL_BLAS} and allows a meaningful reference to the original \adda BiCGStab results (also duplicated here from Fig.~\ref{fig:fig3_dda_runtime_gpu}). Unfortunately, timing inside the OpenCL code has not yet been implemented; thus, the FFT timing is not available for \cmd{OCL_BLAS} mode.

On all three GPUs and for both grid sizes, \cmd{OCL_BLAS} systematically outperforms the default \adda OpenCL mode, except on the NVIDIA RTX~2000~Ada for $n_x=250$, where the computation fails due to an out-of-memory error (limit is around $n_x=230$). The improvement is most pronounced on the H200, where solver time is reduced by factors of approximately 9 ($n_x=150$) and 6.5 ($n_x=250$). This confirms that, in the standard OpenCL mode, performance is limited primarily by CPU--GPU-transfer latency rather than by GPU compute throughput.

The trends observed here closely mirror those of \ifdda: switching from A100 to H200 increases the \cmd{OCL_BLAS} solver performance by a factor of approximately 1.8, consistent with the expected scaling trend based on GPU memory bandwidth.

\bibliographystyle{elsarticle-num} 
\bibliography{references.bib}
\end{document}
\typeout{get arXiv to do 4 passes: Label(s) may have changed. Rerun}